\documentclass[aps,prx,twocolumn,superscriptaddress]{revtex4}
\usepackage{amsmath,amssymb,wasysym}
\usepackage{graphicx}
\usepackage{dcolumn}
\usepackage{bm}
\usepackage{braket}
\usepackage{verbatim}
\usepackage{float}
\usepackage{bbold}
\usepackage{color}
\usepackage{xcolor}
\usepackage{relsize}
\usepackage{amsthm}
\usepackage{enumerate}
\usepackage{soul,xcolor}
\usepackage[normalem]{ulem}
\usepackage[T1]{fontenc}
\usepackage{adjustbox}
\usepackage[colorlinks=true ,urlcolor=blue,urlbordercolor={0 1 1}]{hyperref}

\renewcommand{\>}{\rangle}

\newcommand{\be}{\begin{equation}}
\newcommand{\ba}{\begin{align}}
\newcommand{\ee}{\end{equation}}
\newcommand{\bea}{\begin{eqnarray}}
\newcommand{\eea}{\end{eqnarray}}
\newcommand{\beq}{\begin{equation}}
\newcommand{\eeq}{\end{equation}}
\newcommand{\beqn}{\begin{eqnarray}}
\newcommand{\eeqn}{\end{eqnarray}}

\renewcommand{\hat}[1]{{\widehat #1}}

\begin{document}
\widetext

\title{Variational wavefunction for Mott insulator at finite $U$ using ancilla qubits}
\author{Boran Zhou}
\thanks{These authors contribute equally.}
\affiliation{William H. Miller III Department of Physics and Astronomy, Johns Hopkins University, Baltimore, Maryland, 21218, USA}
\author{Hui-Ke Jin}
\thanks{These authors contribute equally.}
\affiliation{School of Physical Science and Technology, ShanghaiTech University, Shanghai 201210, China}
\author{Ya-Hui Zhang}
\affiliation{William H. Miller III Department of Physics and Astronomy, Johns Hopkins University, Baltimore, Maryland, 21218, USA}

\date{\today}

\begin{abstract}
The Mott regime with finite $U$ offers a promising platform for exploring novel phases of matter, such as quantum spin liquids (QSL) that exhibit fractionalization and emergent gauge field. Here, we provide a new class wavefunction, dubbed ancilla wavefunction, to capture both charge and spin (gauge) fluctuations in QSLs at finite $U$. The ancilla wavefunction can unify the Fermi liquid and Mott insulator phases with a single variation parameter $\Phi$ tuning the charge gap. As $\Phi \rightarrow \infty$, the  wavefunction reduces to the Gutzwiller projected state, while at $\Phi=U/2$, it is effectively equivalent to applying an inverse Schrieffer–Wolff transformation to the Gutzwiller projected state. 
This wavefunction can be numerically simulated in the matrix product state representation, and its performance is supported by numerical results for both one- and two-dimensional Hubbard models.
Besides, we propose the possibility of a narrow regime of fractional Fermi liquid phase between the usual Fermi liquid and the Mott insulator phases close to the metal-insulator transition---a scenario typically overlooked by the conventional slave rotor theory.    
Our ancilla wavefunction offers a novel conceptual framework and a powerful numerical tool for understanding Mott physics.

\end{abstract}

\maketitle

\section{Introduction}

Mott physics is at the center of condensed matter physics, due to its close relation to many important topics including high-temperature superconductor~\cite{lee2006doping}, Mott metal-insulator transitions, and various forms of quantum magnetism~\cite{auerbach2012interacting}. The simplest effective model to describe Mott physics is the fermionic Hubbard model, which is known to be relevant to a variety of condensed matter systems, ranging from conventional solid-state material such as cuprates~\cite{lee2006doping}, to Moir\'e systems~\cite{wu2018hubbard,tang2020simulation,regan2020mott,wang2020correlated,kennes2021moire}, and to optical lattices in cold atom systems~\cite{esslinger2010fermi}. Despite decades of intense study, many unresolved problems remain in this seemingly simple model.

At half-filling ($1/2$ filling per spin), a Mott insulating phase begins to form when the Hubbard $U$ dominates over the hopping $t$. For $U$ approaching infinity, where all charge fluctuations are frozen and only spin degrees of freedom are activated, P.~W.~Anderson proposed the celebrated idea of resonating valence bond state~\cite{Anderson73,anderson1987resonating} --- now recognized as the first example of quantum spin liquids (QSL)~\cite{savary2016quantum,norman2016colloquium,zhou2017quantum,broholm2020quantum}. These novel phases of matter are characterized by their fractionalized excitations, emergent gauge fields, and non-trivial projective symmetry groups (PSG) pattern~\cite{wen2002quantum}. Over the past decades, the Gutzwiller projected parton wavefunction has proven to be a powerful tool for effectively describing QSL states~\cite{anderson1987resonating,anderson2004physics,paramekanti2001projected,capriotti2001resonating,capello2005variational,ran2007projected,becca2017quantum,lee2006doping}. However, extending the Gutzwiller wavefunction to finite $U$ remains challenging and lacks a straightforward approach. 
One choice is to naively generalize the Gutzwiller projector to $P_G=\prod_i (1-\alpha n_{i;\uparrow} n_{i;\downarrow})$, which inevitably leads to an unexpected metallic phase in a Mott phase once $\alpha<1$. 
To capture the finite charge gap in the Mott regime, it may be necessary to incorporate additional factors on top of Gutzwiller projected wavefunctions, such as a doublon-holon binding factor~\cite{guertler2014kagome} or a more complex Jastrow factor of $J_d=e^{-\frac{1}{2} V_{ij} n_i n_j}$~\cite{capello2005variational, capello2006unconventional}, potentially including backflow terms~\cite{tocchio2011backflow}.

However, for all the aforementioned approaches, the Mott gap is not controlled by a single parameter. An even more serious issue arises when the underlying state is a QSL with an emergent gauge flux per unit cell, which can break translation and/or lattice rotation symmetries in a projective manner~\cite{wen2002quantum}. This is generally evident in states such as a Dirac QSL~\cite{wen2002quantum} or certain chiral spin liquids (CSLs)~\cite{song2021doping}. In this case, the Jastrow factor explicitly breaks the physical lattice symmetry because the density fluctuations in $J_d=e^{-\frac{1}{2} V_{ij} n_i n_j}$ incorrectly manifest the non-trivial PSG as a real symmetry breaking pattern~\footnote{There is no such issue in the Gutzwiller projected wavefunction at infinite $U$ because the density fluctuations are completely frozen there.}. Recently, a translational invariant CSL was identified from unbiased density matrix renormalization group (DMRG)~\cite{White1992,White1993} calculations in the finite $U$ regime~\cite{szasz2020chiral}.  The CSL here is identified as a U(1)$_2$ state and should have a $\pi$ flux ansatz in parton construction~\cite{song2021doping}. Nevertheless, in the variational study using the standard Jastrow wavefunctions, this more natural ansatz was excluded~\cite{tocchio2021hubbard} because it breaks the translation symmetry due to the artifacts introduced by the Jastrow factor. Apparently, representing a QSL with non-trivial PSG in the finite $U$ regime is an unresolved fundamental problem. 


In this work, we address this challenge by proposing a conceptual framework to unify the charge fluctuations and emergent non-trivial PSGs in Mott insulators. 
The idea is that, given a QSL state as a Gutzwiller projected state, we introduce a single parameter to embed the fluctuations of doublons and holons into the background of this Gutzwiller projected state (or a general many-body spin state). This aspect has not been extensively explored before and could provide a crucial foundation for understanding metal-insulator transitions and doped Mott insulators --- two of the most significant and challenging problems in condensed matter physics.

We propose a new class of variational wavefunctions, dubbed ancilla wavefunction, by introducing ancilla qubits to mediate the entanglement between emerged partons in QSLs and physical electrons. This ancilla wavefunction was initially proposed by one of the authors from a phenomenological perspective for the pseudogap metal in hole-doped cuprates~\cite{zhang2020pseudogap}. Nevertheless, its validity and qualities within a microscopic Hamiltonian have not been carefully verified. Focusing on half-filling, we provide analytical proof demonstrating that the ancilla wavefunction recovers the classic Gutzwiller projected wavefunction at infinity $U$. For large $U$, we find that it is equivalent to applying the inverse Schrieffer–Wolff transformation to the Gutzwiller projection with a correction in the linear order of $t/U$.
We also represent the ancilla wavefunction as a matrix product state (MPS)~\cite{perezgarcia2007,orus2014,ciracRMP2021}, which allows us to numerically examine its performance in the whole regime of $U/t$. With the analytical argument in a generic dimension and numerical benchmarks in one-dimensional (1D) and 2D systems, we conjecture that the wavefunction remains valid throughout the entire Mott insulating regime, extending down to the zero charge gap limit on a generic lattice. 

With a simple extension to finite doping, the ancilla wavefunction can describe a fractional Fermi liquid (FL*) phase that violates the Luttinger theorem without breaking any symmetry.  This state may be a strong candidate for the mysterious pseudogap metal observed in hole-doped cuprate~\cite{zhang2020pseudogap}. Notably, we further point out that such a FL* phase can in principle exist in a narrow region of $U/t$ between the Fermi liquid and the Mott insulator phase, even at half-filling. This scenario has not been well established before, as conventional approaches such as slave rotor theory~\cite{slave_rotor,senthil2008theory} fail to capture it. We propose to investigate such an exotic state in future numerical studies. 

The rest of the paper is organized as follows. In Section~\ref{sec:setup}, we present the basic setup for the ancilla wavefunction and describe the procedure for converting it into MPS form. The physical properties of the ancilla wavefunction are detailed in Section~\ref{sec:anccila}, where we show that it reproduces the Gutzwiller projection in the limit of infinity $U$ and is equivalent to the inverse Schrieffer-Wolff transformation at large $U$. We also explore the physical meaning of the introduced ancilla fermions in this section. In Section~\ref{sec:1dhubbard} and Section~\ref{sec:KHmodel}, we numerically demonstrate the efficiency of the ancilla wavefunction for both the 1D Hubbard chain and 2D Kitaev-Hubbard model, respectively. Focusing on the emergent topological degenerate and dressed Wilson loop in the Kitaev-Hubbard model, we highlight that the QSL state at finite $U$ can be efficiently captured by the ancilla wavefunction. The intermediate FL* phase around the metal-insulator transition is pointed out in Section~\ref{sec:pseudogap}. Finally, we summarize in Section~\ref{sec:summary}.

\begin{figure}[!t]
\includegraphics[width=1.0\linewidth]{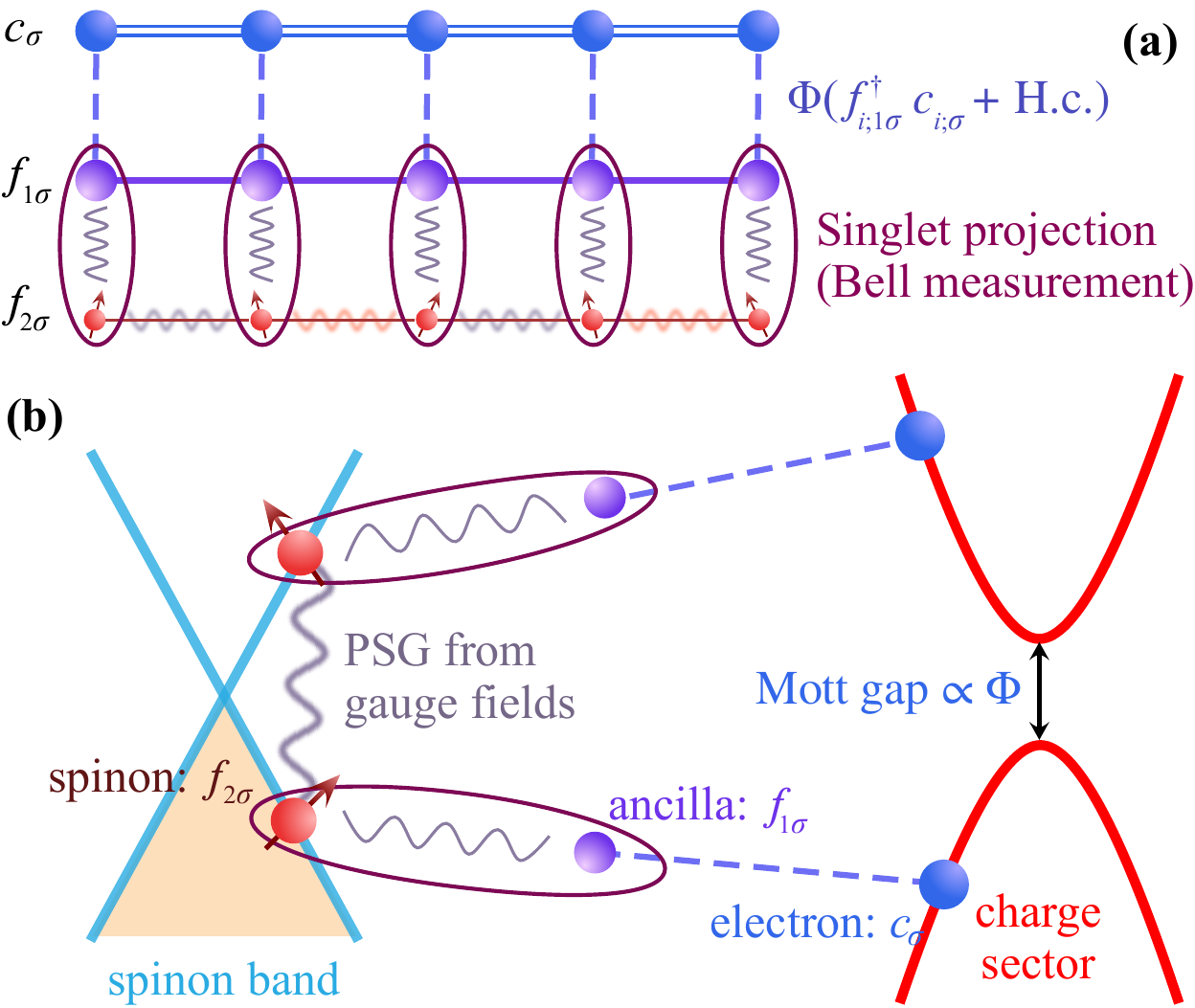}
\caption{{\bf The illustration of the ancilla wavefunctions.} (a) At each lattice site $i$, we have one physical electron $c_{i;\sigma}$ (blue balls) and two spinful ancilla fermions, $f_{i;1\sigma}$ (purple balls) and $f_{i;2\sigma}$ (red balls with arrows), respectively. The only entanglement between $c_{\sigma}$ and the first ancilla $f_{1\sigma}$ is introduced by the hybridization $\Phi$ term; see Eq.~\eqref{eq:HMe}. The entanglement between $f_{1\sigma}$ and $f_{2\sigma}$ is established by the singlet projection which requires two ancilla fermions at the same lattice site to form a spin singlet.
This projection is equivalent to the Bell measurement with post-selection (or feedback) and implements a quantum teleportation. (b) The ancilla fermions $f_{2\sigma}$ can be considered as Abrikosov fermions (spinons) with an emergent SU(2) gauge field (the wavy lines) that supports non-trivial PSG patterns. The non-trivial entangled state of $f_{2\sigma}$ can be ``transmitted'' to the physical $c_{\sigma}$ fermion through the spin-singlet projection, namely, quantum teleportation.
The ancilla wavefunction can preserve all the lattice symmetries even though the PSG patterns of $f_{2\sigma}$ may break some of the symmetries in the mean-field level. 
The hybridization of $c_{\sigma}$ and $f_{1\sigma}$ gives the upper and lower Hubbard bands.
The Mott gap of the Hubbard model in Eq.~\eqref{eq:hubbard} at the large $U$ limit is generally proportional to the hybridization parameter $\Phi$. When $\Phi\rightarrow\infty$, the ancilla wavefunctions can exactly resolve the Gutzwiller-projected wavefunctions, corresponding to the infinite $U$ limit.}~\label{Fig3}
\end{figure}

\section{Basic setup}~\label{sec:setup}


We start with the classic fermionic Hubbard model:
\begin{equation}
    H=-t\sum_{\langle ij \rangle,\sigma}\left(c^\dagger_{i;\sigma} c_{j;\sigma}+\mathrm{H.c.}\right)+U\sum_i n_{i;\uparrow}n_{i;\downarrow},~\label{eq:hubbard}
\end{equation}
where $c_{i;\sigma}$ $(\sigma=\uparrow,\downarrow)$ denotes the annihilation operator for physical electrons and $n_{i;\sigma}=c^\dagger_{i;\sigma} c_{i;\sigma}$ is a spin-$\sigma$ density operator at lattice site $i$.

Our goal is to introduce the ancilla wavefunction as an efficient trial state for the Hubbard model~Eq.~\eqref{eq:hubbard}. The approach of this ancilla wavefunction is illustrated in Fig.~\ref{Fig3} and elaborated on in detail below. 
At each lattice site $i$, we introduce two spinful ancilla fermions, $f_{i;1\sigma}$ and $f_{i;2\sigma}$.  The ancilla wavefunction is then expressed as 
$$\ket{\Psi_c}=P_S \ket{\Psi_0[c,f_1,f_2]},$$ 
where $\ket{\Psi_c}$ is the state in the physical Hilbert space spanned by electrons $c_{i;\sigma}$ only, and $\ket{\Psi_0}$ belongs to the enlarged Hilbert space spanned by $c_{\sigma}$ and the ancilla fermions $f_{1\sigma}$ as well as $f_{2\sigma}$. The singlet projection operator $P_S$ enforces that two local ancilla fermions, $f_{i;1\sigma}$ and $f_{i;2\sigma}$, form a spin singlet at each lattice site $i$. 

Here, we restrict ourselves to a very specific ansatz, given by 
\begin{equation}
    \ket{\Psi_0}=\ket{\text{Slater}[c,f_1]}\otimes\ket{\text{Gauss}[f_2]}.~\label{eq:psi0}
\end{equation} 
The above form assumes that before the singlet projection $P_S$, the physical electron couples only to $f_{1\sigma}$. Meanwhile, $f_{2\sigma}$ is treated as an independent state, decoupled from the other fermions. The first part, $\ket{\text{Slater}[c,f_1]}$, is a Slater determinant determined by the free fermion Hamiltonian 
\begin{align}~\label{eq:HMe}
    H^M_{e}=& - t_c \sum_{\langle ij \rangle,\sigma}\left( c^\dagger_{i;\sigma} c_{j;\sigma} +\mathrm{H.c.}\right)- \mu_c \sum_{i,\sigma} c^\dagger_{i;\sigma} c_{i;\sigma}\notag \\
    &+ t_1 \sum_{\langle ij \rangle,\sigma }\left(f^\dagger_{i;1\sigma} f_{j;1\sigma} +\mathrm{H.c.}\right)-\mu_1 \sum_{i,\sigma} f^\dagger_{i;1\sigma} f_{i;1\sigma} \notag \\ 
    &+\Phi \sum_{i,\sigma} \left(c^\dagger_{i;\sigma} f_{i;1\sigma}+\mathrm{H.c.}\right),
\end{align}
in which the average density of $f_{1\sigma}$ is fixed at one particle per site by the chemical potential $\mu_1$. The average density of the physical electron $c_\sigma$ can be adjusted to any value by tuning $\mu_c$, but for now we mainly focus on the half filling case.  $\Phi$ is the only term that entangles electrons $c_{\sigma}$ and the ancilla fermions $f_{1\sigma}$, making it the most crucial variational parameter in the model. 
Later we will show that at half-filling, $\Phi$ is proportional to the charge gap of the Hubbard model~\eqref{eq:hubbard} in the large $U$ regime.
The second part, $\ket{\text{Gauss}[f_2]}$, represents a ground state of a parton mean-field theory of the ancilla $f_{2\sigma}$, which usually can be characterized by a non-trivial PSG~\cite{wen2002quantum}. As will be explained below, $f_{2\sigma}$ should be considered as the familiar Abrikosov fermions (spinons), manifesting an effective SU(2) gauge structure. 
Meanwhile, the singlet projection $P_S$ automatically imposes a single-occupancy constraint on the $f_{2\sigma}$ fermions in $\ket{\text{Gauss}[f_2]}$.
Therefore, $\ket{\text{Gauss}[f_2]}$ in Eq.~\eqref{eq:psi0} can be replaced by its Gutzwiller-projected format. Indeed, any physical spin-1/2 wavefunction in practice can be used in place of $\ket{\text{Gauss}[f_2]}$.

Given $\ket{\text{Slater}[c,f_1]}$ and $\ket{\text{Gauss}[f_2]}$ as ingredients, the ancilla wavefunction is constructed by projecting $f_{1\sigma}$ and $f_{2\sigma}$ out to form a product of a spin singlet: $$\ket{s}=\prod_i \frac{1}{\sqrt{2}} (f^\dagger_{i;1\uparrow}f^\dagger_{i;2\downarrow}-f^\dagger_{i;1\downarrow}f^\dagger_{i;2\uparrow})\ket{0}.$$
Then, the ancilla wavefunction is expressed as a form purely in the physical Hilbert space: 
$$\ket{\Psi_c}= \sum_{c} (\braket{c,s|\Psi_0}) \ket{c},$$ 
where $\ket{c}$ is summed over the many-body states spanned by the physical electrons $c_{\sigma}$ with a fixed particle number (e.g., haff-filling).  

\begin{figure}[!t]
\includegraphics[width=1.0\linewidth]{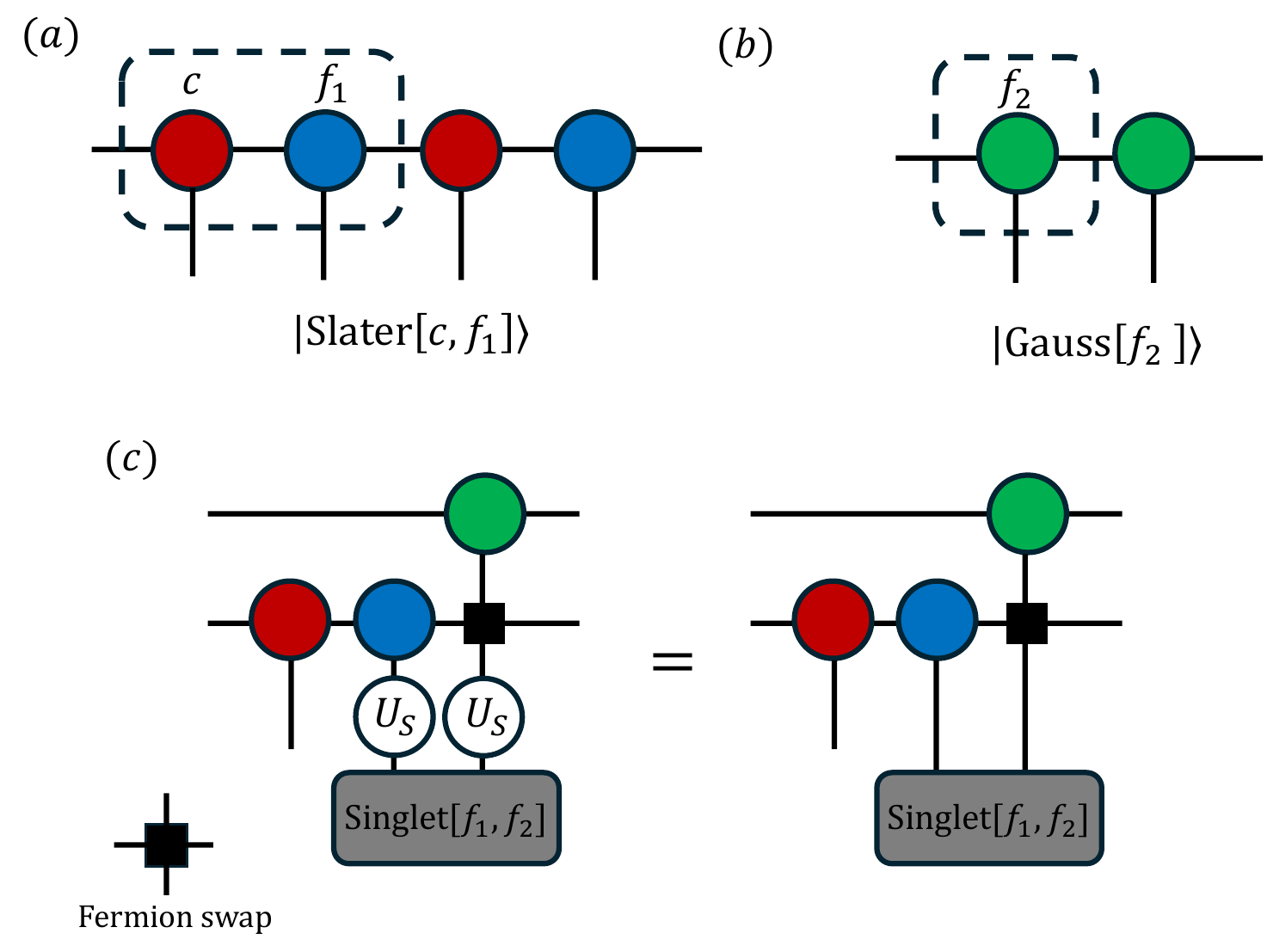}
\caption{{\bf MPS representation for the ancilla wavefunction} (a-b) MPS representations for (a) $\ket{\mathrm{Slater}\left[c,f_1\right] }$ and (b) $\ket{\mathrm{Gauss}\left[f_2\right]}$. (c) Illustration of singlet projection in one unit cell. The fermion swap is a diagonal tensor to correctly handle the fermionic anticommutation relations~\cite{PhysRevA.81.010303,PhysRevB.81.165104}. $U_S$ denotes an local SU(2) transformation that preserves the ancilla wavefunction.} 
\label{Fig1}
\end{figure}

Though $f_{1\sigma}$ and $f_{2\sigma}$ disappear after projection $P_S$, their essential quantum fluctuations and entanglement structures are inherited by the physical electrons $c_{\sigma}$. 
The effect of ancilla fermions and the projection procedure can be easily understood using the framework of MPS~\cite{orus2014practical,ciracRMP2021}; see Fig.~\ref{Fig1}. 
In the MPS representation, the vertical leg at each site corresponds to the physical degrees of freedom, and the horizontal legs between two neighboring sites are associated with $\chi$ virtual states, where $\chi$ is known as the bond dimension of the MPS.  In our approach, we first construct a MPS for $\ket{\text{Slater}[c,f_1]}$ based on $H^M_{e}$, as shown in Fig.~\ref{Fig1}(a).  And then we convert a parton mean-field ansatz into a MPS as $\ket{\text{Gauss}[f_2]}$; see Fig.~\ref{Fig1}(b). To obtain $\ket{\Psi_0}$ in Eq.~\eqref{eq:psi0}, we combine these two MPSs through a tensor product, as illustrated in Fig.~\ref{Fig1}(c).
The singlet projection is carried out by contracting the legs of $f_{i;1\sigma}$ and $f_{i;2\sigma}$ for each site $i$. After this contraction, the resulting MPS retains only the physical legs corresponding to the electrons $c_{i\sigma}$ and meanwhile, acquires the entanglement of $\ket{{\rm Gauss}[f_2]}$ via $f_{1\sigma}$. Eventually, we obtain a variational wavefunction for the Hubbard model in Eq.~\eqref{eq:hubbard}.


We would like to stress that the ancilla wavefunction is invariant under a SU(2)$_1\otimes$SU(2)$_2\otimes$SU(2)$_S$ gauge transformation~\cite{zhang2020pseudogap}. Note that the singlet projection $P_S$ indeed imposes three constraints on the ancilla fermions at each site $i$: (i) $n_{i;1}=1$; (ii) $n_{i;2}=1$; and (iii) $\vec S_{i;1}+\vec S_{i;2}=0$, where $n_{i;a}$ and $\vec S_{i;a}$ is the density and spin operators of the ancilla $f_{a\sigma}$ ($a=1,2$), respectively. 
The single-occupancy constraint for $f_{a\sigma}$ can be fixed as a U(1)$_a$ gauge field. Moreover, due to additional particle-hole transformation~\cite{wen2002quantum} in spin-$1/2$ fermions, each U(1) gauge field is enlarged to SU(2)$_{a}$, which leaves the spin operators $\vec S_{i;a}$ unchanged. Thus, they are just the SU(2) gauge structure found in the standard parton construction theory, which emerges in the context of fractionalized $S=1/2$ spins~\cite{wen2002quantum}.
The third SU(2) gauge structure can straightforwardly manifest in the MPS representation, as shown in Fig.~\ref{Fig1}(c). 
The singlet projector $P_S$ acts as an operator with the bra and ket states being two spin bases formed by $f_{1\sigma}$ and $f_{2\sigma}$, respectively. Because $P_S$ itself is a spin singlet, it is invariant under arbitrary local spin SU(2) transformation: $$P_S(i)=U_{S}^\dagger{}(i)P_S(i)U_{S}(i),$$ where $P_S=\prod{}P_S(i)$ and $U_S(i)$ is a arbitrary local SU(2) transformation at site $i$. Strictly speaking, the SU(2) matrices $U_S(i)$ on the left and right sides act on the local states of $f_{i;1\sigma}$ and $f_{i;2\sigma}$, respectively; see Fig.~\ref{Fig1}(c). This SU(2)$_1\otimes$SU(2)$_2\otimes$SU(2)$_S$ gauge structure indicates that the entanglement of the non-trivial PSG pattern in $f_{2\sigma}$, which emerges from one of the SU(2)$_a$ gauge fields, can be transmitted into physical electrons through the spin-singlet projection. This process effectively acts as quantum teleportation, as discussed in the next section.

\section{Quantum teleportation interpretation}~\label{sec:anccila}

We first focus on the special case with $\Phi=+\infty$ at half-filling. Then, the wavefunction $\ket{\mathrm{Slater}[c,f_{1}]}$ can be exactly expressed as $$\ket{\mathrm{Slater}\left[c,f_1\right]}=\prod_i\left(\frac{1}{2}(c^\dagger_{i;\uparrow}-f^\dagger_{i;1\uparrow})(c^\dagger_{i;\downarrow}-f^\dagger_{i;1\downarrow})\right)\ket{0}.$$ 
Note that at each site $i$, $c_{i;\sigma}$ and $f_{i;1\sigma}$ form a spin singlet, which indeed is a perfect Einstein–Podolsky–Rosen (EPR) pair.  The projection $P_S$ thus can be understood as a Bell measurement with post selection. As in the classic quantum information context~\cite{nielsen2002quantum}, this projection (Bell measurement) implements quantum teleportation and teleports the quantum entanglement in $\ket{{\rm Gauss}[f_2]}$ to the physical electrons $c_{\sigma}$.  Moreover, after simple calculations, one can prove that (see Appendix~\ref{app:infU}) 
\begin{equation} 
    \ket{\Psi_c(\Phi\rightarrow\infty)}=P_G \ket{\text{Gauss}_{f_2}\left[c\right]},
\end{equation}
with $P_G$ the familiar Gutzwiller projection $\prod_i (1- n_{i;\uparrow} n_{i;\downarrow})$ to enforce the single occupancy of electron for each site $i$.  The Gaussian state $\ket{\text{Gauss}_{f_2}\left[c\right]}$ in terms of $c_{\sigma}$ is the same as the input parton ansatz $\ket{\text{Gauss}\left[f_2\right]}$ in terms of $f_{2\sigma}$. Naturally, the ancilla wavefunction serves as a good trial wavefunction in the infinite $U$ limit.

\subsection{Equivalance to inverse Schrieffer-Wolff transformation at large U} 

At $U\rightarrow\infty$, the exact single-occupancy constraint on the physical electron $c_{\sigma}$ is indirectly imposed through its coupling to $f_{1\sigma}$ via the $\Phi\rightarrow\infty$ term. For a large but finite $U$, this constraint can be softened by reducing $\Phi$ from its infinitely large value, wherein the charge fluctuations are incorporated. Note that in the regime with a Mott gap, there is flexibility in choosing $t_c$ and $t_1$ without changing the final wavefunction, as long as $t_c + t_1$ remains constant; see Appendix~\ref{app:largeU}.
Here, we simply set $t_c = t$ and $t_1 = 0$, leading to $\Phi$ as the only remained variational parameter in Eq.~\eqref{eq:HMe}.

With a finite $\Phi$, the electrons $c_{\sigma}$ and ancilla $f_{1\sigma}$ remain entangled, though they no longer form a perfect EPR pair at each site. Consequently, the projection results in an imperfect quantum teleportation. At the linear order of $t/U$, the correction can be worked out analytically:
\begin{equation}
    \left|\Psi_c\left(\Phi=\frac{U}{2}\right)\right\rangle\approx \exp[-\mathrm{i}S] \ket{\Psi_c(\Phi=+\infty)}\label{eq:SW}
\end{equation}
where $\exp[-\mathrm{i}S]$ is the inverse of the Schrieffer-Wolff transformation in the standard $t/U$ expansion for the Hubbard model~\cite{macdonald1988t}. Therefore, in the large $U$ regime, our ancilla wavefunction accurately captures the charge fluctuations with $\Phi = \frac{U}{2}$. It indicates that at large $U$, one can treat $\Phi$ as a fixed parameter. The only remaining free parameters to be optimized are the parton ansatz in terms of the $f_{2\sigma}$ fermions. The details of the Schrieffer-Wolff transformation are provided in Appendix~\ref{app:largeU}.

The choice of spin-part parton ansatz $\ket{{\rm Gauss}[f_{2}]}$ depends on the specific problem being addressed.
Nevertheless, this is a well-explored problem within the context of spin models (e.g., see  Refs.~\cite{anderson1987resonating, anderson2004physics, paramekanti2001projected, capriotti2001resonating, capello2005variational, ran2007projected, becca2017quantum}). For a specific Hubbard model with charge fluctuation, one can employ corresponding well-established spin ansatz for $\ket{{\rm Gauss}[f_{2}]}$ and optimize them accordingly within our formalism.

The efficiency of our ancilla wavefunction can be demonstrated by presenting a wavefunction for the CSL in the triangular lattice Hubbard model in the intermediate $U$ regime (see Appendix~\ref{app:jastrow}).
This CSL was recently identified as the ground state by DMRG~\cite{szasz2020chiral}, and it is proposed to exhibit a $\pi$ flux per unit cell. However, conventional Jastrow factor approaches tend to yield a state breaking translation symmetry, which must be energetically less favorable~\cite{tocchio2021hubbard}. In contrast, our ancilla wavefunction maintains lattice symmetries even when $\ket{\mathrm{Gauss}[f_{2}]}$ is a $\pi$-flux ansatz, and thereby can gain more energy. 

\subsection{Physical meaning of the ancilla fermions}   

This section provides more intuitive interpretations of the ancilla fermions $f_{1\sigma}$ and $f_{2\sigma}$.  
By noting that the ancilla wavefunction has the structure $\ket{\Psi_c}=P_S \ket{\Psi_0}$, one can apply $f_{1\sigma}$ and $f_{2\sigma}$ operators to $\ket{\Psi_0}$ to obtain variational excited states. When $\Phi\rightarrow\infty$, as detailed in Appendix~\ref{app:infU}, we find that 
\begin{equation*}
\begin{split}
    &P_Sf^\dagger_{i;1\sigma} \ket{\Psi_0}=c^\dagger_{i;\sigma} \ket{\Psi_c}, \\ 
    &P_Sf^\dagger_{i;2\sigma} \ket{\Psi_0}=0,\\
    &P_Sf^\dagger_{i;2\sigma} f_{i;2\sigma'}\ket{\Psi_0}=c^\dagger_{i;\sigma}c_{i;\sigma'} \ket{\Psi_c}.
\end{split}
\end{equation*} 
Therefore, the action of $f_{1\sigma}^\dagger$ is the same as that of electron operator $c_{\sigma}^\dagger$, while $f_{2\sigma}^\dagger $ alone is meaningless.  Nevertheless, $f_{i;2\sigma}^\dagger f_{i;2\sigma}$ is meaningful and behaves as a physical spin operator. 
One natural interpretation is that $f_{1\sigma}$ effectively acts as the electron operator $c_{\sigma}$, while $f_{2\sigma}$ can be viewed as a neutral spinon operator, particularly when $\Phi$ is large.

A better way to understand this is to utilize the aforementioned SU(2)$_1\otimes$SU(2)$_2\otimes$SU(2)$_S$ gauge theory to resolve the single projection. 
We denote the first two SU(2) gauge fields as $a$ and $b$, which couple to the ancilla fields $f_{1\sigma}$ and $f_{2\sigma}$, respectively.
And the third SU(2)$_S$ gauge, which couples to both the $f_{1\sigma}$ and $f_{2\sigma}$ fermions, is referred to as $\alpha$. 
The physical electron $c_{\sigma}$ is gauge invariant as expected.  
With a large $\Phi$ in $H^{M}_e$ in Eq.~\eqref{eq:HMe}, the gauge fields $a$ and $\alpha$ are both Higgsed as they couple to $f_{1\sigma}$~\cite{zhang2020deconfined}. Hence $a$ and $\alpha$ can be ignored as they are in the Higgsed phase of a gauge theory.
It indicates that the ancilla $f_{1\sigma}$ can now be treated as the physical electron $c_{\sigma}$~\footnote{More precisely, the gauge field $a$ is locked to the external probing field $A$, then $f_{1\sigma}$ also couples to the physical electric and magnetic field as an electron.}. In contrast, the gauge field $b$ is untouched by the term $\Phi$. The coupling of $f_{2\sigma}$ to the gauge field $b$ mirrors the familiar Abrikosov fermion description of QSLs. As mentioned in Section~\ref{sec:setup}, $f_{2\sigma}$ can be identified as the standard Abrikosov fermionic spinon~\cite{wen2002quantum}, which enables us to recover any QSLs within the framework of ancilla wavefunction.
Moreover, after applying the singlet projection $P_S$, the nontrivial PSG pattern in the gauge field $b$ will not be interpreted as a physical symmetry-breaking, ensuring that the ancilla wavefunction naturally preserves all the lattice symmetries.

In summary, with a finite $\Phi$, we have two separate sectors: a charge sector formed by $c_{\sigma}$ and $f_{1\sigma}$, and a spin sector formed by the spinon $f_{2\sigma}$ as shown in Fig.\ref{Fig3}(b). 
The charge gap is approximately $\Delta_c = 2\Phi$. It is natural to expect that the ancilla wavefunction can be extended further into the regime with a relatively small charge gap by simply decreasing the value of $\Phi$. Below we confirm this conjecture for 1D systems with explicit numerical results.

\begin{figure}[!t]
\includegraphics[width=1.0\linewidth]{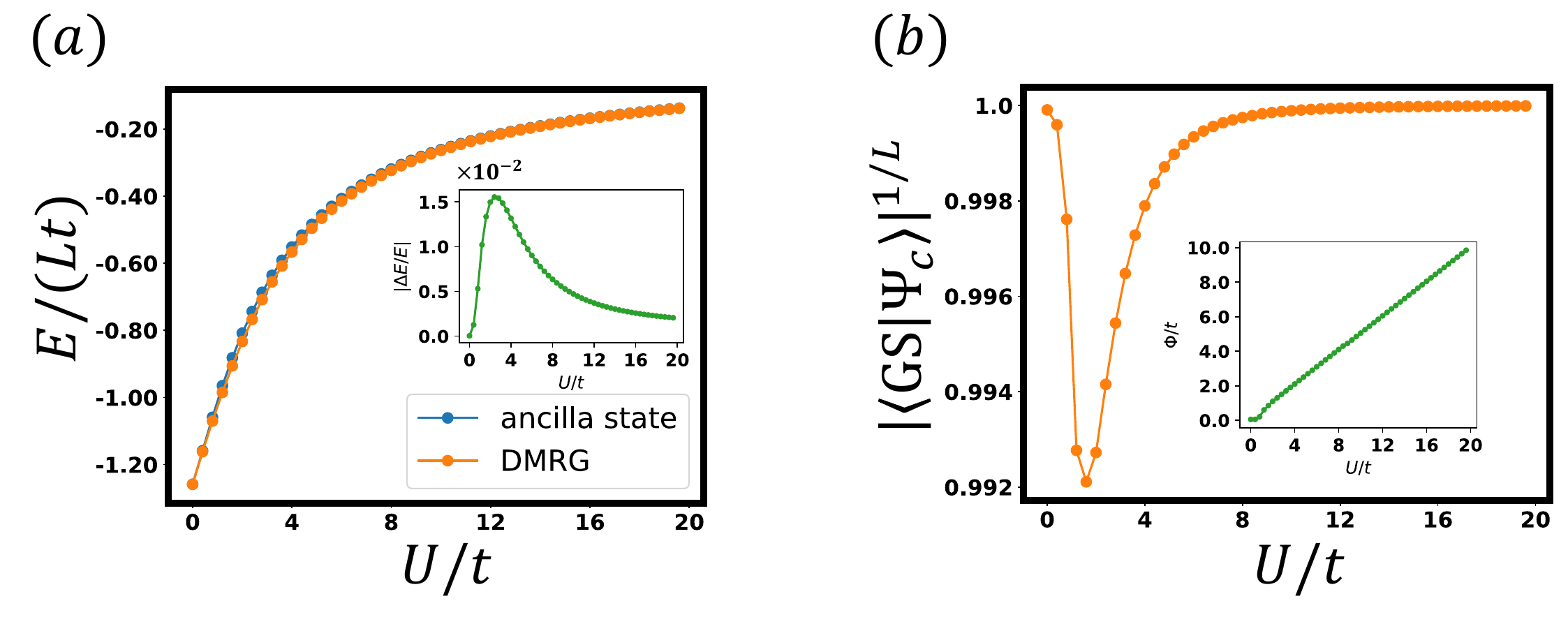}
\caption{{\bf Benchmark of the ancilla wavefunction for the 1D Hubbard chain with $L=50$.}
(a) The per-site energy of the ancilla wavefunction $\ket{\Psi_c}$ ($E_\mathrm{ancilla}$) is compared with the DMRG results ($E_\mathrm{DMRG}$) for various $U/t$. Inset: The relative energy difference $|(E_\mathrm{ancilla}-E_\mathrm{DMRG})/E_\mathrm{DMRG}|$.
(b) The per-site overlap between the ancilla wavefunction and the ground state obtained by DMRG versus $U/t$. In the regime $U/t>8$, the per-site overlap exceeds $0.9997$, leading to an overall system overlap greater than $0.987$. Inset: The optimized parameter $\Phi/t$ is proportional to $U/t$ as $\Phi\sim U/2$, which is consistent with the analytical predictions in Section~\ref{sec:anccila}. }
\label{Fig4}
\end{figure}

\section{Applications}

In practice, we construct the ancilla wavefunction in MPS representation with the help of the TeNPy Library~\cite{tenpy}. 
The recently developed methods based on MPS~\cite{PhysRevB.92.075132,Tu2020,Gabriel2021,Jin2020,aghaei2020efficient,Jin2021,Jin2022} enable us to prepare $|\mathrm{Slater}[c,f_1]\rangle$ and $|\mathrm{Gauss}[f_2]\rangle$ on (quasi-)1D geometries. For systems with spin SU(2) symmetry, one only needs to calculate the spin-up channel of $|\mathrm{Slater}[c,f_1]\rangle$, as the spin-down channel is an identical replica. Additionally, the step of constructing $|\mathrm{Gauss}[f_2]\rangle$ from a parton ansatz can be replaced by directly simulating the ground state of the corresponding spin model using DMRG. 
The Gutzwiller zipper method~\cite{aghaei2020efficient} is employed to perform a tensor product of these three states. Following this, we implement the singlet projection $P_S$ to convert the ancilla wavefunction $\ket{\Psi_c}$ into an MPS form.
Note that the calculations can be further simplified at half-filling. In this particular case, one can apply Gutzwiller projection $P_G$ to $|\mathrm{Gauss}[f_2]\rangle$ before implementing the tensor product.

\subsection{Application to the 1D Hubbard chain}~\label{sec:1dhubbard}

We first adopt the 1D Hubbard chain to benchmark the performance of our ancilla wavefunction. Focusing on the half-filling limit, the Hamiltonian for the model is given by:
\begin{equation}
    H_{1D} = -t\sum_{i,\sigma}(c^\dagger_{i;\sigma}c_{i+1;\sigma}+\mathrm{H.c.})+U\sum_{i}n_{i;\uparrow}n_{i;\downarrow}.
\end{equation}
At infinite $U$, the above model is reduced to the classic 1D antiferromagnetic Heisenberg chain. To improve the accuracy, our ansatz for the $f_{2\sigma}$ part is chosen as the ground state of the 1D Heisenberg chain. We emphasize that similar results can be obtained using the Gutzwiller projected parton ansatz $P_G|\mathrm{Gauss}[f_2]\rangle$, as demonstrated in Appendix~\ref{app:results}.

For a given $U/t$, we optimize $\Phi$ to minimize the energy of  $|\Psi_c\rangle$ for the Hubbard chain, as shown in Fig.~\ref{Fig4}(a). Remarkably, the optimized parameter $\Phi$, which results in the best variational energy, scales as $\Phi\sim U/2$, as shown in the inset of Fig.~\ref{Fig4}(b). This behavior is consistent with the analytical arguments mentioned in Section~\ref{sec:anccila}.

To assess the quality of the wavefunction, we further calculate the overlap between $\ket{\Psi_c}$ and $\ket{\mathrm{GS}}$ the ground state for the Hubbard chain obtained by DMRG. The results, presented in Fig.\ref{Fig4}(b), show that the fidelity (the per-site overlap) exceeds 0.99 across the entire range of $U/t$, demonstrating the notable performance of our ancilla wavefunction with just one variational parameter $\Phi$. In particular, the fidelity surpasses 0.9997 when $U/t > 8$.

\subsection{Application to the 2D Kitaev-Hubbard model}~\label{sec:KHmodel}

\begin{figure}
    \includegraphics[width=1.\linewidth]{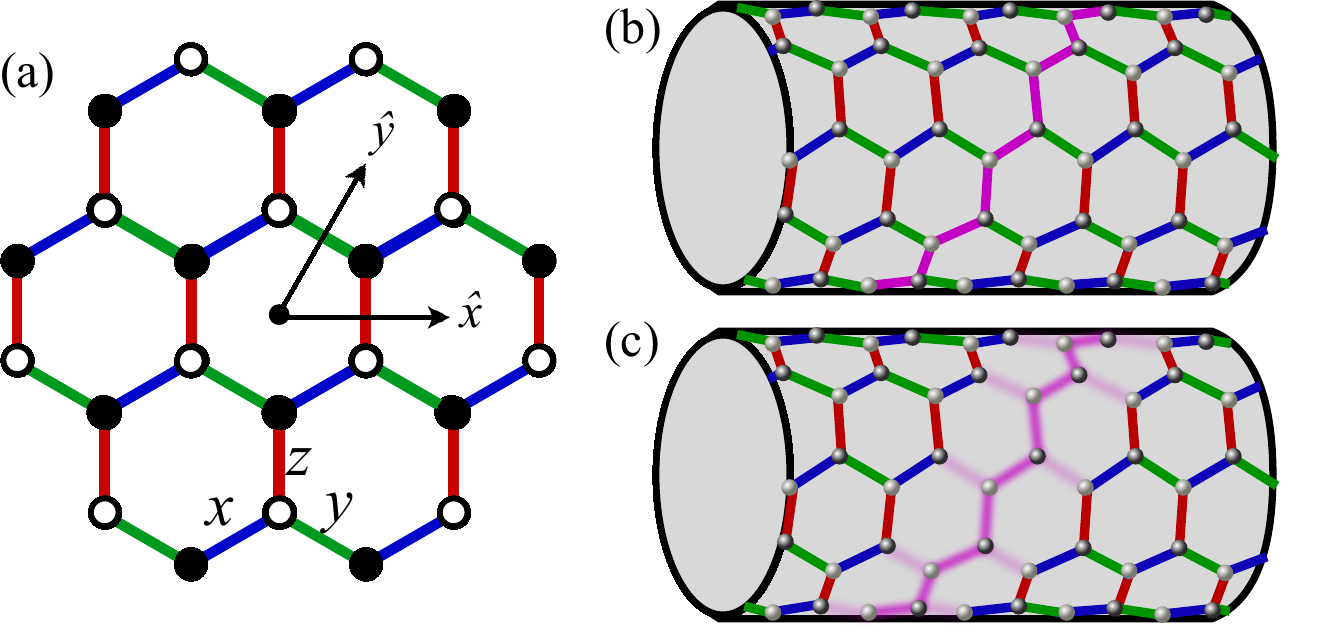}
    \caption{The Kitaev-Hubbard model and emergent Wilson loop. (a) The sketch for the Hamilotnian in Eqs.~\eqref{eq:KHmodel} and \eqref{eq:KSmodel}. The blue, green, and red bonds denote the $x$, $y$, and $z$ types of bonds, respectively. (b) The Kitaev-Hubbard model on a cylindrical geometry, in which the $\hat{x}$ boundary is open while the $\hat{y}$ boundary is periodic. The exact Wilson loop is defined on the purple zigzag line as a closed loop through the cylinder. (c) The emergent Wilson loop is defined on the dressed purple zigzag loop. }\label{fig:KHmodel}
\end{figure}

In this section, we demonstrate the efficacy of the ancilla wavefunction by applying it to investigate a 2D Hubbard model, referred to as the Kitaev-Hubbard model. The Hamiltonian for the Kitaev-Hubbard model, defined on the honeycomb lattice, is given by~\cite{Duan2003,Hassan2013,Faye2014,Liang2014} (Einstein notation used for repeated spin indices)
\begin{equation}
H_{K}=-\sum_{\langle{}ij\rangle{}\in{}\alpha}\left(t^{\sigma\sigma'}_{\alpha}c^\dagger_{i;\sigma}c^{}_{j;\sigma'}+{\rm H.c.}\right)+U\sum_{i}n^{}_{i;\uparrow}n^{}_{i;\downarrow},\label{eq:KHmodel}
\end{equation}
where hopping amplitude $t^{\sigma\sigma'}_\alpha\equiv{}(\delta_{\sigma\sigma'}+\tau^\alpha_{\sigma\sigma'})t/2$ $(\alpha=x,y,z)$, $\tau^\alpha$ are three Pauli matrices, and
$\langle{}il\rangle{}\in{}\alpha$ run all the $\alpha$-type NN bonds on the honeycomb lattice; see Fig.~\ref{fig:KHmodel}(a). Without Hubbard interactions, Eq.~\eqref{eq:KHmodel} gives rise to a free band structure of semimetal with Dirac-type excitations at half filling.  

Although the ground state of the Kitaev-Hubbard model generally remains unknown when $U>0$, with an infinitely large Hubbard $U$, this model reduces to the exactly-solvable Kitaev honeycomb model~\cite{Kitaev06}, given by pure spin degrees of freedom as 
\begin{equation}
H_{Kitaev}=\sum_{\langle{}ij\rangle{}\in{}\alpha}JS^\alpha_{i}S^\alpha_{j},\quad{}J=t^2/U. \label{eq:KSmodel}
\end{equation}
Note that here the physical spin $S^\alpha_{i}$, formed by electron operators $c_{i;\sigma}$, should be distinguished from $S^\alpha_{i;a}$, the spin operators of the ancilla fermions $f_{i;a\sigma}$ ($a=1,2$). Nevertheless, to find the exact solution, one must adopt a four-Majorana representation for physical spins, by which Eq.~\eqref{eq:KSmodel} is mapped into a free Majorana fermion theory coupled to a static $Z_2$ gauge field. 
This four Majorana representation is equivalent to the familiar Abrikosov fermion representation in terms of $f_{2\sigma}$, up to a SU(2)$_2$ gauge rotation~\cite{Nayak2011}. Therefore, the ground state of Eq.~\eqref{eq:KSmodel}, known as Kitaev QSL, is classified as a Z$_2$-type QSL with a nontrivial PSG~\cite{You2012}. 

This exact solution of the Kitaev honeycomb model sheds light on understanding the effective low-energy physics for the Kitaev-Hubbard model at finite Hubbard $U$. When expressed in terms of Abrikosov fermion $f_{2\sigma}$, the Kitaev QSL becomes a Gutzwiller projected BCS state which can be efficiently represented in the form of MPS on cylindrical geometries~\cite{Jin2021}. This representation facilitates the construction of ancilla wavefunctions for the Kitaev-Hubbard model. 

In practice, we embed the honeycomb lattice on a finite cylinder with $L_y$ and $L_x$ unit cells along the periodic $y$-direction and open $x$-direction, respectively; see Fig.~\ref{fig:KHmodel}. The total number of lattice sites is thus $N=2L_xL_y$. We use bond dimensions of $\chi_{1}$ and $\chi_{2}$ to prepare $|{\rm Slater}[c, f_{1}]\>$ and $|{\rm Gauss}[f_{2}]\>$, respectively, leading to the final MPS of ancilla wavefunction with a bond dimension of $\chi=\chi_{1}\times{}\chi_{2}$. For preparing $|{\rm Slater}[c,f_{1}]\>$, the ancilla fermions $f_{i;1\sigma}$ are fixed to be dispersionless and only locally hybridized to the electron $c_{i;\sigma}$ via a variational parameter $\Phi$. 

\begin{figure}[!t]
    \includegraphics[width=1.\linewidth]{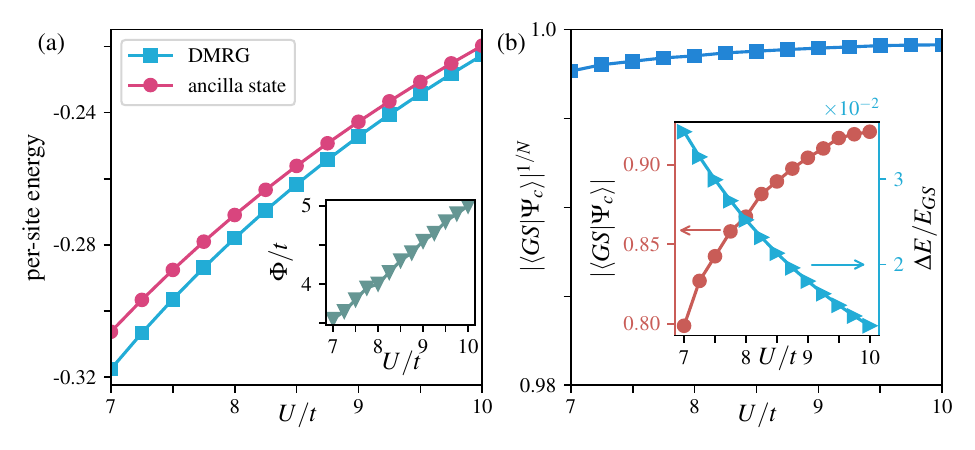}
    \caption{ The ancilla wavefunction as a trial state for the Kitaev-Hubbard model. (a) The variational energy $E_{KH}$ and DMRG energy as functions of $U/t$. Inset: the optimized parameter $\Phi/t$, which gives rise to the best $E_{KH}$, as a function of $U/t$. (b) The fidelity (per-site overlap) between ancilla wavefunction and DMRG state $|{\rm GS}\>$ as a function of $U/t$. Inset: (Left axis) the corresponding overlap for the total system. (Right axis) the relative energy difference. 
    The results are obtained on a cylinder with $L_x=12$ and $L_y=4$ ($N$=96). The bond dimension for DMRG is 8000. }\label{fig:FigKHeng}
\end{figure}

The quality of this ancilla wavefunction is evaluated by examining the variational energy $E_{KH}(\Phi)$ for Eq.~\eqref{eq:KHmodel}.  
As predicted by our analysis, the best variational energy is obtained at $\Phi\approx U/2$; see the inset of Fig.~\ref{fig:FigKHeng}(a). Since the charge fluctuations primarily contribute to the variational energies, it is not surprising that $E_{KH}$ is more sensitive to $\chi_{1}$ rather than $\chi_{2}$. Therefore, we examine the convergence of the variational energy $E_{KH}$ as the bond dimension $\chi_1$ increases, with detailed results in Appendix~\ref{app:results}.  
In Fig.~\ref{fig:FigKHeng}(a), we show the optimized variational energy $E_{KH}$ and the DMRG energy as functions of $U/t$. The relative difference between the extrapolation value of $E_{KH}$ at $\chi_1\rightarrow{}\infty$ and the energy obtained by DMRG is about $2\%$. In the regime of $U/t>9$, the per-site wave function fidelity between ancilla wavefunction and DMRG state $|{\rm GS}\>$ is about $\approx0.9989$, resulting in overlap for the total system being of $\approx{}0.90$ on a cylinder with $L_x=12$ and $L_y=4$ ($N$=96); see Fig.~\ref{fig:FigKHeng}(b).  
We emphasize that our ancilla wavefunction here involves only a single variational parameter $\Phi$. Despite this simplicity, its performance is already impressive. Moreover, its performance could potentially be further enhanced by introducing additional parameters.

We would like to emphasize that the ancilla wavefunction demonstrates a QSL state with finite charge fluctuations, which are unavoidably present in realistic materials. To reveal this QSL nature, we investigate the emergent Wilson loop in the Kitaev-Hubbard model. 

In the Kitaev honeycomb model, one can define Wilson loop operators wrapping around the cylinder, e.g., $$W_y=-\prod_{i\in{}\mathbb{L}}(2S^y_{i}),$$ where $\mathbb{L}$ is a closed zigzag loop demonstrated in Fig.~\ref{fig:KHmodel}(b). This Wilson loop operator, whose eigenvalues are $Z_2$ number, commutes with the Kitaev honeycomb model $H_{Kitaev}$. It indicates that the Kitaev QSL exhibits two-fold topologically degenerate ground states on cylinders, 
which are characterized by the expectation value of $W_y$ as $w_y\equiv{}\langle{}W_y\rangle=\pm{}1$. 

However, such a Wilson loop is not necessarily evident for the Kitaev-Hubbard model, because $W_y$ does not commute with $H_{K}$ in Eq.~\eqref{eq:KHmodel}. Nevertheless, one still can prepare two Kitaev QSL states as $|{\rm Gauss}[f_{2}]\>$ in both $w_y=\pm{}1$ sectors, and use them to construct their respective ancilla wavefunctions. The expectation values of the Wilson loop operator for these two ancilla wavefunctions are approximately $w_y\approx\pm{}0.85$ at $\Phi/t=4.5$ on cylinders with $L_y=4$. We further perform DMRG simulations initialized with these two ancilla wavefunctions at $U/t=2\Phi/t=9$, and find that the values and signs of $w_y$ are preserved. It indicates that the MPS stays in the respective $Z_2$ sector during DMRG optimizations, and the Kitaev-Hubbard model exhibits {\rm emergent} topological degeneracy in its Mott phase. 

By noting the physical meaning of the ancilla wavefunctions, one can conclude that this {\em emergent} topological degeneracy in the Kitaev-Hubbard model is characterized by the dressed Wilson loop operators, $$\tilde{W}_y=\exp(-iS){}W_y\exp(iS),$$ where $\exp(iS)$ is the corresponding Schrieffer-Wolff transformation in Eq.~\eqref{eq:SW}. In the large $U$ limit, the Schrieffer-Wolff transformation can be approximated as $\exp(iS)\approx1+iS$ since $iS\propto{}t/U\ll{}1$. Moreover, the generator $iS$ is defined on the NN bonds only, analogous to the kinetic energy of doublons and holons. Clearly, the dressed operator $\tilde{W}_y$ is mainly defined on the zigzag loop $\mathbb{L}$ as well as on the dangling NN bonds connecting to $\mathbb{L}$, as illustrated in Fig.~\ref{fig:KHmodel}(c). Therefore, $\tilde{W}_y$ remains spatially localized around $\mathbb{L}$, maintaining its nature as a thicker but still proper string operator. We evaluate the expectation value of $\langle{}\tilde{W}_{y}\rangle\approx\langle{}W_{y}+[-iS, W_y]\rangle\approx{}\pm{}1.04$ for $U/t=9$. Its amplitude is slightly larger than one because the higher-order corrections $\sim{}O(t^2/U)$ are omitted.

\section{Intermediate pseudogap metal}~\label{sec:pseudogap} 

The wavefunction here can be easily extended to the case of finite doping by simply tuning the chemical potential $\mu_c$ for electrons $c_{\sigma}$. In this case, our ancilla wavefunction can describe a symmetric pseudogap metal dubbed as fractional Fermi liquid (FL*)  phase~\cite{zhang2020pseudogap,mascot2022electronic,christos2023model}. In this section, we point out that a similar FL* phase is possible around the metal-insulator transition even at half-filling.

\begin{figure}[!t]
\includegraphics[width=1.0\linewidth]{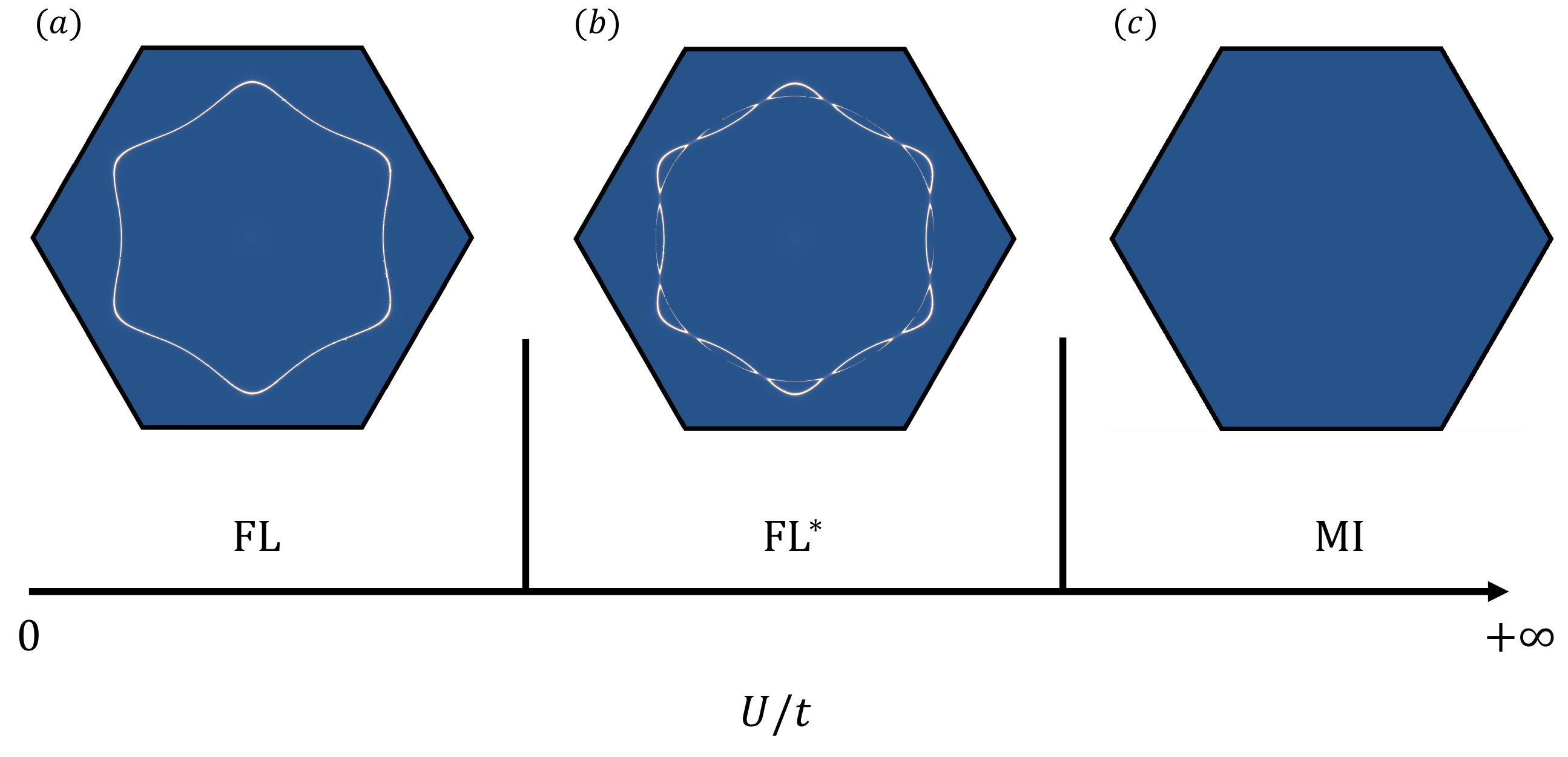}
\caption{A conjectured phase diagram tuned by $U/t$ at half-filling on the triangular lattice. The left critical point is associated with the onset of $\Phi$, while the right critical point is simply a Lifshitz transition.  FL, FL*, and MI label the Fermi liquid, fractional Fermi liquid, and Mott insulator. Across these two transitions, we assume that $f_{2\sigma}$ forms a CSL parton state which is illustrated in Appendix~\ref{app:jastrow}. (a)-(c) The contour plots of the electron spectral weight $A_c(\mathbf{k},\omega=0)$.  (a) For $\Phi=0$, the system is a Fermi liquid. (b) For $\Phi=0.1$, the system is a FL* phase with equal size of electron and hole pockets and dominated spectral weight in disconnected Fermi arcs. (c) For large $\Phi$, the system is Mott insulator with $A_c(\mathbf{k},0)=0$. }
~\label{Fig5}
\end{figure}

To demonstrate the FL* phase for the undoped system, we consider the $t_1$-$t_2$ Hubbard model on the triangular lattice. Accordingly, the Hamiltonian $H^M_e$ in Eq.~\eqref{eq:HMe} consists of NN hopping $t_{c,1}$ and next nearest neighbor hopping $t_{c,2}$ for the electrons $c_{\sigma}$. For simplicity, we assume that the ancilla fermions $f_{1\sigma}$ only exhibit NN hopping $t_1$. Without loss of generality, we focus on the case of $t_{c,1}=t_{1}=1$ and  $t_{c,2}=-0.82$.

At half-filling, we have shown that $\Phi\sim U/2$ in the large U regime efficiently describes a Mott insulator in our ancilla wavefunction.
For $\Phi=0$, the electrons $c_{\sigma}$ are decoupled to the ancilla $f_{1\sigma}$ and then $\ket{\Psi_{c}}$ is the Fermi sea ground state for the kinetic part of the Hubbard model, which corresponds to a Fermi liquid state; see Fig.~\ref{Fig5}(a).
Therefore, the bandwidth-controlled metal-insulator transition should naturally correspond to the onset of $\Phi$ in the ancilla wavefunction. However, generically, there may be a small mismatch between the Fermi surfaces of $c_{\sigma}$ and $f_{1\sigma}$ (except for the 1D system) without fine-tuning. In this case, a small $\Phi$ can not fully gap out all the Fermi surfaces, leading to an intermediate pseudogap metal phase (FL* phase)  for relatively small $\Phi$. 
To demonstrate this, we study the electron spectral weight function~\cite{coleman2015introduction}
\begin{equation}
    A_c(\mathbf{k},\omega)=\frac{1}{\pi}\mathrm{Im}G_c(\mathbf{k},\omega-\mathrm{i}0^+),
\end{equation}
where $G_c(\mathbf{k},\omega)$ is the Green's function of electron $c_{\sigma}$.; see details in Appendix~\ref{app:spectral}. 
 As illustrated in Fig.~\ref{Fig5}(b), one can observe equal-sized electron and hole pockets in $A_c(\mathbf{k},\omega=0)$, with the backside originating from $f_{1\sigma}$ and thus having a smaller spectral weight. Consequently, disconnected Fermi arcs, which are similar to those seen in underdoped cuprates, may appear at finite temperatures.  
As $\Phi$ increases with rising $U/t$, these pockets shrink and eventually, the system evolves into a Mott insulator with $A_c(\mathbf{k},0)=0$. The possibility of an intermediate FL* phase was overlooked by previous slave rotor approaches~\cite{slave_rotor}. On the other hand, our wavefunction provides a legitimate framework for this scenario, and it is interesting to carry out detailed numerical studies to explore this exotic phase further in the future. 


\section{Discussion and Conclusion}~\label{sec:summary}

In summary, we proposed an ancilla wavefunction for Mott insulators with a finite charge gap. We analytically show that this wavefunction resolves the Gutzwiller projected wavefunction when $U\rightarrow{}\infty$. For large $U$, this wavefunction is equivalent to applying an inverse Schiffer-Wolff transformation to the Gutzwiller projected wavefunction, with a correction in the linear order of $t/U$. 
We conjecture that the ancilla wavefunction also works well in the weak Mott regime, as supported by our benchmark results for the 1D Hubbard chain at half-filling. We also show that the extension of the wavefunction can capture a symmetric pseudogap metal (a FL* phase) in the one-orbital model, which is indeed a challenge for the previous approaches.
Meanwhile, the ancilla wavefunction exhibits the flexibility to capture the QSLs with both charge fluctuations and non-trivial PSG patterns. We demonstrate it by studying the Kitaev-Hubbard model on the honeycomb lattice, in which emergent topological degeneracy and dressed Wilson loop are explored. Therefore, we anticipate the ancilla wavefunction will be useful for unifying the weak and strong Mott regimes.

Our work raises a whole range of questions for the future
research. First, it will be interesting to develop higher-dimension tensor network states or variational Monte Carlo algorithms~\cite{shackleton2024,muller2024polaronic} to deal with the 2D Hubbard model with both SU(2) and more general SU(N) spins. Second, one could explore the wavefunction upon doping with application to high $T_c$ superconductivity in underdoped cuprates. Third, searching for the FL* phase as the ground state of some concrete models will be worthwhile. 

In conclusion, our ancilla wavefunction offers a novel theoretical framework and a powerful numerical tool for understanding Mott physics ranging from weak Mott insulators to pseudogap metals, and to QSL as well as high $T_c$ superconductors.

\section{Acknowledgement}
We thank Subir Sachdev and Henry Shackleton for helpful discussions. We thank Zhehao Dai for the discussions on the tensor network algorithm.  This work was supported by the
National Science Foundation under Grant No. DMR-2237031. The work by YHZ was performed in part at Aspen Center for Physics, which is supported by National Science Foundation grant PHY-2210452. The numerical simulation was carried out at the Advanced Research Computing at Hopkins (ARCH) core facility (rockfish.jhu.edu), which is supported by the National Science Foundation (NSF) grant number OAC 1920103.
\bibliographystyle{apsrev4-1}
\bibliography{main}

\appendix 

\section{Jastrow and ancilla wavefunction at finite U with non-trivial PSG}~\label{app:jastrow}

In this section, we compare the Jastrow wavefunction with our ancilla wavefunction in the regime with a finite $U$.  We will argue that the Jastrow factor approach fails to represent any spin liquid state (QSL) with a non-trivial projective symmetry group (PSG) pattern.

A general form of the Jastrow factor wavefunction is:
\begin{equation}
    \ket{\Psi_{J}}=e^{- \frac{1}{2} \sum_{ij}V_{ij} n_i n_j} |\text{Slater}[c]\rangle.
    \label{eq:jastrow}
\end{equation}
In practice, one usually applies the above Jastrow factor to suppress the charge fluctuations in a Gaussian parton ansatz.
If one uses the Gutzwiller projection $P_G=\prod_i (1-n_{i;\uparrow}n_{i;\downarrow})$ instead of Eq.~\eqref{eq:jastrow}, the obtained state is a pure spin state without any charge fluctuations. However, if we use the Jastrow factor, the fermions in the Gaussian parton ansatz now carry both spin and charge degrees of freedom.  With a sufficiently singular $V_{ij}$, the charge gap is open and Eq.~\eqref{eq:jastrow} describes a Mott insulator of QSL. Nevertheless, the density-density correlation is influenced by the QSL which breaks lattice symmetries in a projective manner.   On the other hand, in a Mott insulator, we should expect spin-charge separation: the suppression of the density fluctuation happens at the energy scale of $\Delta_c \sim U$, while the QSL part is decided by the energy scale of $J \sim \frac{4 t^2}{U}\ll\Delta_c$. Therefore we should expect that the density fluctuations are roughly independent of the QSL ansatz. The Jastrow factor wavefunction fails on this aspect.

\begin{figure}[ht]
\includegraphics[width=0.7\linewidth]{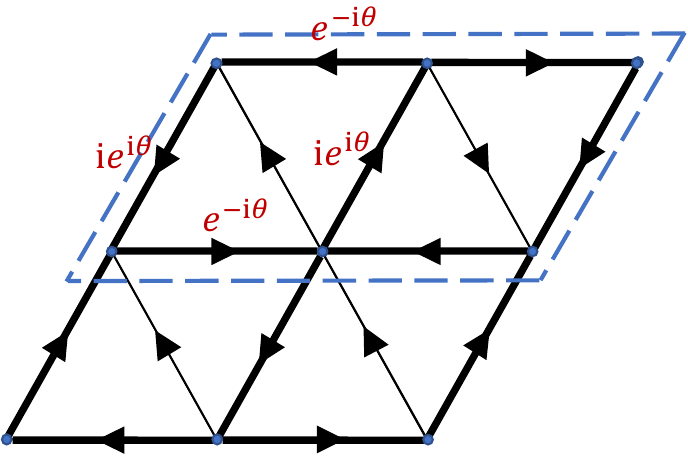}
\caption{A chiral spin liquid (CSL) ansatz on triangular lattice for fermionic spinons ($f_{2\sigma}$). The blue dashed line indicates the doubled unit cell. The orientation of the arrow along the bond represents the phase of the hopping parameter in this ansatz, resulting in a $\pi$ flux per unit cell. In the Jastrow factor approach, the charge fluctuations fell the doubled unit cell and the final state explicitly breaks translation symmetry. In contrast, the ancilla wavefunction preserves all the lattice symmetries after singlet projection because the density fluctuation of $f_{2\sigma}$ is completely frozen as in the familiar Gutzwiller projected wavefunction.}
\label{fig:csl_ansatz}
\end{figure}

To further reveal the problem, we consider a chiral spin liquid ansatz on the triangular lattice.   We have fermionic spinons ($f_{2\sigma}$) in an ansatz with a doubled unit cell and a flux $\pi$ per unit cell, as illustrated in Fig.~\ref{fig:csl_ansatz}. Now the translation symmetry along the $x$ direction is broken in the mean-field level. After Gutzwiller projection, the resulting state is still translation invariant~\cite{wen2002quantum}. 

For finite $U$, the Jastrow factor wavefunction incorrectly treats the fractionalized spinons ($f_{2\sigma}$) as physical electrons $c_{\sigma}$, which causes the unphysical SU(2) gauge fluctuations in the parton ansatz to manifest as physical fluctuations. For instance, we  calculate the expectation value of the bond current operator
\begin{equation}
    \langle T_{ij} \rangle=\sum_{\sigma}\langle ic^\dagger_{i\sigma} c_{j\sigma} -i c^\dagger_{j\sigma} c_{i\sigma} \rangle.
\end{equation}
For an infinite $U$, one can easily verify that $\langle T_{ij} \rangle = 0$. However, as long as the Hubbard $U$ is finite, the nonzero $\langle T_{ij} \rangle$ must follow the same PSG pattern as the mean field ansatz in Fig.~\ref{fig:csl_ansatz},  breaking the translation symmetry. The Jastrow factor suppresses the density fluctuations, but can not alter this property qualitatively. As a result, the final wavefunction is not translationally invariant and thus is a quite poor ansatz for QSL with nontrivial PSG.  The issue with the Jastrow wavefunction originates from the artifact that both charge and spin are represented by the same fermion. In a Mott insulator, however, we expect a separation between the charge sector and the spin sector, which the Jastrow wavefunction fails to capture properly.

On the other hand, our ancilla wavefunction can easily represent a translationally invariant CSL with an arbitrary charge gap. 
The singlet projection $P_S$ enforces that there is only one $f_{2\sigma}$ particle per site, the same as the familiar Gutzwiller projection for the spin model in the limit of infinite $U$.Therefore, there are no density fluctuations corresponding to $f_{2\sigma}$ fermions since 
$f_{2\sigma}$ are purely charge neutral as the Abrikosov fermion. Meanwhile, the charge sector is separately controlled by $\Phi$ hybridizing electrons $c_{\sigma}$ and ancilla $f_{1\sigma}$, which is independent of the parton ansatz in terms of $f_{2\sigma}$.

\section{Ancilla wavefunction of Mott insulator at $\Phi=+\infty$}~\label{app:infU}

In the limit of $\Phi\rightarrow+\infty$, the hopping terms of $c_{\sigma}$ and $f_{1\sigma}$ can be ignored. Therefore, $c_{\sigma}$ and $f_{1\sigma}$ form spin-singlet at each site $i$:
\begin{equation}\label{eqGS1}
    \ket{\mathrm{Slater}[c,f_1]}=\prod_{i=1}^N\left(\frac{1}{2}(c^\dagger_{i;\uparrow}-f^\dagger_{i;1\uparrow})(c^\dagger_{i;\downarrow}-f^\dagger_{i;1\downarrow})\right)\ket{0}.
\end{equation}
Here we assume a generic lattice on arbitrary dimension. 
In the main text we define that $$\ket{\Psi_0}=\ket{\mathrm{Slater}[c,f_1]}\otimes\ket{\mathrm{Gauss}[f_2]}.$$ Here for convenience, we replace $\ket{\mathrm{Gauss}[f_2]}$ by its Gutzwillr projected form as 
\begin{equation}\label{eqGS2}
    \ket{\Psi_2}=P_G\ket{\mathrm{Gauss}[f_2]}=\sum_{\{\sigma_i\}}\psi(\{\sigma_i\}) \prod_{i=1}^Nf^\dagger_{i;2\sigma_i}\ket{0},
\end{equation}
Then, without changing any results, we can rewrite $\ket{\Psi_0}$ as
\begin{equation}\label{eqGS}
\ket{\Psi_0}=\ket{\mathrm{Slater}[c,f_1]}\otimes\ket{\Psi_2},
\end{equation}
which still satisfies $\ket{\Psi_c}=P_S\ket{\Psi_0}=\sum_c\langle c,s|\Psi_0\rangle|c\rangle$. 

By combining Eq.~\eqref{eqGS1}, Eq.~\eqref{eqGS2} and Eq.~\eqref{eqGS}, the final expression of $\ket{\Psi_0}$ is written as:
\begin{widetext}
\begin{equation}~\label{eqGSform}
\begin{split}
    \ket{\Psi_0}=&\sum_{\{\sigma_i\}}\psi(\{\sigma_i\})\prod_{i=1}^N\left(\frac{1}{2}(c^\dagger_{i;\uparrow}-f^\dagger_{i;1\uparrow})(c^\dagger_{i;\downarrow}-f^\dagger_{i;1\downarrow})f^\dagger_{i;2\sigma_i}\right)\ket{0}\\
    =&\sum_{\{\sigma_i\}}\psi(\{\sigma_i\})\prod_{i=1}^N\left(\frac{1}{2}(-\sigma_ic^\dagger_{i;\sigma_i}f^\dagger_{i;1\bar{\sigma}_i}f^\dagger_{i;2\sigma_i}+\mathrm{other terms})\right)\ket{0},
\end{split}
\end{equation}
where $\sigma_i=1,-1$ for spin up and down, respectively. The projection operator $P_S=\sum_c |c\rangle\langle c,s|$, which enforces $f_1$ and $f_2$ to form spin singlet at each site $i$, can be written as:
\begin{equation}
   P_S=\prod_{i=1}^N \left(\frac{1}{\sqrt{2}}(1-n_{i;1\uparrow})(1-n_{i;1\downarrow})(f_{i;2\downarrow}f_{i;1\uparrow}-f_{i;2\uparrow}f_{i;1\downarrow})\right)=\prod_{i=1}^N P_{S,i},
\label{EqProjection}
\end{equation}
where we define $P_{S,i}=\frac{1}{\sqrt{2}}(1-n_{i;1\uparrow})(1-n_{i;1\downarrow})(f_{i;2\downarrow}f_{i;1\uparrow}-f_{i;2\uparrow}f_{i;1\downarrow})$. Substituting Eq.~\eqref{EqProjection} into Eq.~\eqref{eqGSform} and multiplying a normalization factor, we obtain that
\begin{equation}\label{eqGSphiinfty}
    \ket{\Psi_c}=\sum_{\{\sigma_i\}}\psi(\{\sigma_i\})\prod_{i=1}^N c^\dagger_{i;\sigma_i}\ket{0}.
\end{equation}
It is clear to see that the form of Eq.\eqref{eqGSphiinfty} is the same as that of Eq.~\eqref{eqGS2}, indicating that our ancilla wavefunction resolves Gutzwiller projected state when $\Phi\rightarrow\infty$.

Based on Eq.~\eqref{eqGSform}, we examine how fermion operators affect the final projected state. The basic two cases are simply applying one $c_{\sigma}/c^\dagger_{\sigma}$ or one $f_{1\sigma}/f_{1\sigma}^\dagger$ on top of $|\Psi_0\>$ before implementing the singlet projection. For the electron operators, it is obvious that
\begin{equation}
    c_{j;\sigma}\ket{\Psi_0}=\sum_{\{\sigma_i\}}\psi(\{\sigma_i\})(-1)^{\sum_{k<j}n_k}\prod_{i=1}^N\left(\frac{1}{2}(\delta_{ij}c_{j;\sigma}+1-\delta_{ij})(c^\dagger_{i;\uparrow}-f^\dagger_{i;1\uparrow})(c^\dagger_{i;\downarrow}-f^\dagger_{i;1\downarrow})f^\dagger_{i;2\sigma_i}\right)\ket{0},
\end{equation}
then we have,
\begin{equation}
    \begin{split}
        P_Sc_{j;\sigma}\ket{\Psi_0}=&(-1)^{j-1}\sum_{\{\sigma_i\}}\psi(\{\sigma_i\})\prod_{i=1}^N (\delta_{ij}c_{j;\sigma}+1-\delta_{ij})c^\dagger_{i;\sigma_i}\ket{0}
        =c_{j;\sigma}\ket{\Psi_c}=c_{j;\sigma}P_S\ket{\Psi_0}.
    \end{split}
\end{equation}
As for the ancilla $f_{1\sigma}$, we find that 
\begin{equation}
\begin{split}
    f_{j;1\sigma}\ket{\Psi_0}=&\sum_{\{\sigma_i\}}\psi(\{\sigma_i\})(-1)^{\sum_{k<j}n_{k}}\prod_{i=1}^N\left(\frac{1}{2}(\delta_{ij}f_{j;1\sigma}+1-\delta_{ij})(c^\dagger_{i;\uparrow}-f^\dagger_{i;1\uparrow})(c^\dagger_{i;\downarrow}-f^\dagger_{i;1\downarrow})f^\dagger_{i;2\sigma_i}\right)\ket{0}\\
    =&\sum_{\{\sigma_i\}}\psi(\{\sigma_i\})(-1)^{\sum_{k<j}n_{k}}\prod_{i=1}^N\frac{1}{2}\left(\left(-(1-\delta_{ij})c^\dagger_{i;\sigma_i}+\delta_{ij}\delta_{\sigma\sigma_i}\right)\sigma_if^\dagger_{i;1\bar{\sigma}_i}f^\dagger_{i;2\sigma_i}+\mathrm{other \ terms}\right)\ket{0}.
\end{split}
\end{equation}
After implementing projection and renormalization, one can obtain
\begin{equation}
\begin{split}
    P_Sf_{j;1\sigma}\ket{\Psi_0}=&\sum_{\{\sigma_i\}}\psi(\{\sigma_i\})(-1)^{j-1}\prod_{i=1}^N\left((1-\delta_{ij})c^\dagger_{i;\sigma_i}-\delta_{ij}\delta_{\sigma\sigma_i}\right)\ket{0}=-c_{j;\sigma}\ket{\Psi_c}.
\end{split}
\end{equation}
\end{widetext}
Carrying out similar calculations, we prove that
\begin{equation}
    \begin{split}
        P_S c^\dagger_{j;\sigma}\ket{\Psi_0}=&c^\dagger_{j;\sigma}\ket{\Psi_c},\\
        P_S f^\dagger_{j;1\sigma}\ket{\Psi_0}=&c^\dagger_{j;\sigma}\ket{\Psi_c}.
    \end{split}
\end{equation}
Since the singlet projection is a product of local projector as $P_S=\prod_{i=N}^1 P_{S,i}$, the above results can be generalized as:
\begin{equation}
    \begin{split}
        P_{S,j} c_{j;\sigma}\ket{\Psi_0}=&c_{j;\sigma} P_{S,j}\ket{\Psi_0},\\
        P_{S,j} c^\dagger_{j;\sigma}\ket{\Psi_0}=&c^\dagger_{j;\sigma} P_j\ket{\Psi_0},\\
        P_{S,j} f_{j;1\sigma}\ket{\Psi_0}=&-f_{j;1\sigma} P_{S,j}\ket{\Psi_0},\\
        P_{S,j} f^\dagger_{j;1\sigma}\ket{\Psi_0}=&f^\dagger_{j;1\sigma} P_{S,j}\ket{\Psi_0}.
    \end{split}
\end{equation}

For the case with more than one operator, one can sort the order of operators with the site index increasing:
\begin{equation}
    O=(-1)^F\prod_{i=1}^N O_i,
\end{equation}
where $(-1)^F$ comes from the exchange of fermionic operators. The projected state is:
\begin{equation}
\begin{split}
   P_SO \ket{\Psi_0}=&(-1)^F (P_{S,N}P_{S,N-1}...P_{S,1})(O_1O_2...O_N)\ket{\Psi_0}\\
   =&(-1)^F (P_{S,1} O_1)(P_{S,2} O_2)...(P_{S,N} O_N)\ket{\Psi_0}.
\end{split}
\end{equation}
The effect of the operator $P_{S,i}O_i$ is independent of the subsequent operators $(P_{S,i+1}O_{i+1})(P_{S,i+2}O_{i+2})...(P_{S,N}O_N)$. 
For example, we conclude that 
\begin{equation}
\begin{split}
    &P_{S,i} c_{i;\sigma}(P_{S,i+1}O_{i+1})(P_{S,i+2}O_{i+2})...(P_{S,N}O_N)\ket{\Psi_0}\\=&c_{i;\sigma}P_{S,i}(P_{S,i+1}O_{i+1})(P_{S,i+2}O_{i+2})...(P_{S,N}O_N)\ket{\Psi_0}.    
\end{split}
\end{equation}
Therefore, we can prove the following relation:
\begin{equation}\label{eqcdf}
\begin{split}
    P_Sc^\dagger_{i;\sigma}f_{j;1\sigma}\ket{\Psi_0}=&-c^\dagger_{i;\sigma}c_{j;1\sigma}\ket{\Psi_c},\\    P_Sf^\dagger_{i;1\sigma}c_{j;\sigma}\ket{\Psi_0}=&c^\dagger_{i;1\sigma}c_{j;\sigma}\ket{\Psi_c}.
\end{split}
\end{equation}

Because the operators $f_{i;2\sigma}/f_{i;2\sigma}^\dagger$ change the occupancy of $f_{2\sigma}$, we can easily conclude that 
\begin{equation}
    \begin{split}
        P_{S,i}f_{i;2\sigma}\ket{\Psi_0}=&0,\\
        P_{S,i}f^\dagger_{i;2\sigma}\ket{\Psi_0}=&0.
    \end{split}
\end{equation}
For the case with two operators on the same site $i$, we have:
\begin{equation}
\begin{split}
&P_Sf^\dagger_{i;2\sigma}f_{i;2\sigma^\prime}\ket{\Psi_0}\\=&P_S\ket{\mathrm{Slater}[c,f_1]}f^\dagger_{i;2\sigma}f_{i;2\sigma^\prime}\ket{\Psi_2}\\
=&P_S\ket{\mathrm{Slater}[c,f_1]}f^\dagger_{i;2\sigma}f_{i;2\sigma^\prime}\sum_{\{\sigma_j\}}\psi(\{\sigma_j\}) \prod_{j=1}^Nf^\dagger_{j;2\sigma_j}\ket{0}.
\end{split}
\end{equation}
By noting that $f^\dagger_{i;2\sigma}f_{i;2\sigma^\prime}$ only changes $\psi(\{\sigma_j\})$, we find that 
\begin{equation}
P_Sf^\dagger_{i;2\sigma}f_{i;2\sigma^\prime}\ket{\Psi_0}=c^\dagger_{i;\sigma}c_{i;\sigma^\prime}\ket{\Psi_c}.
\end{equation}

\section{Ancilla wavefunction in Hubbard model at large U}~\label{app:largeU}

In this section, we will utilize the inverse of the Schrieffer-Wolff transformation to represent the ground state of the half-filled Hubbard model in the large $U$ limit. Then we prove that this Schrieffer-Wolff transformation is effectively equivalent to our ancilla wavefunction. 

The generic fermionic Hubbard model consists of the kinetic term and interaction term as
\begin{equation}~\label{eqHubbard}
\begin{split}
&H_{\mathrm{Hbbrd}}=T+V,    \\
&T = -\sum_{i,j,\sigma}t_{ij}c^\dagger_{i;\sigma}c_{j;\sigma},~~V = U\sum_i n_{i;\uparrow}n_{i;\downarrow}.
\end{split}
\end{equation}
According to their effect to double-occupied state, the kinetic term $T$ can be further divided into three parts~\cite{macdonald1988t}
\begin{equation*}
    \begin{split}
        &T=T_0+T_1+T_{-1},\\
        &T_0=-\sum_{i,j,\sigma}t_{ij}\left(n_{i;\bar{\sigma}}c^\dagger_{i;\sigma}c_{j;\sigma}n_{j;\bar{\sigma}}+h_{i;\bar{\sigma}}c^\dagger_{i;\sigma}c_{j;\sigma}h_{j;\bar{\sigma}}\right),\\
        &T_1=-\sum_{i,j,\sigma}t_{ij}n_{i;\bar{\sigma}}c^\dagger_{i;\sigma}c_{j;\sigma}h_{j;\bar{\sigma}},\\
        &T_{-1}=-\sum_{i,j,\sigma}t_{ij}h_{i;\bar{\sigma}}c^\dagger_{i;\sigma}c_{j;\sigma}n_{j;\bar{\sigma}},
    \end{split}
\end{equation*}
where $h_{i;\sigma}=1-n_{i;\sigma}$. Here, $T_0$ doesn't change the number of the double-occupied sites, $T_1$ and $T_{-1}$ increase and decrease the number of the double-occupied sites, respectively.
In the large $U$ limit, the double/zero occupancy at each site is forbidden, and the effective Hamiltonian is~\cite{macdonald1988t}
\begin{equation}\label{eqH0}
\begin{split}
    H_{\mathrm{Hbbrd}}^0=&T_0+V+\\&U^{-1}\left([T_1,T_{-1}]+[T_0,T_{-1}]+[T_1,T_0]\right).    
\end{split}
\end{equation}
One can perform a unitary transformation on top of the effective Hamiltonian $H_{\mathrm{Hbbrd}}^0$ and approximate the Hubbard model $H_{\mathrm{Hbbrd}}$ as~\cite{macdonald1988t}
\begin{equation}
        H^\prime=e^{-\mathrm{i}S} H_{\mathrm{Hbbrd}}^0 e^{\mathrm{i}S}=V+T_0+T_1+T_{-1}+O(U^{-2}),
\end{equation}
in which $\mathrm{i}S=U^{-1}(T_1-T_{-1})$. $H^\prime$ is the same as $H_{\mathrm{Hbbrd}}$ with a $O(U^{-1})$ order correction. Straightforwardly, the ground state for  Eq.~\eqref{eqHubbard} can be obtained by performing the unitary transformation $e^{-\mathrm{i}S}$ on the ground state of the pure spin model in Eq.~\eqref{eqH0}, such that 
\begin{equation}\label{eqGSHubbard}
\begin{split}
    \ket{\mathrm{GS},\mathrm{Hubbard}}=&e^{-\mathrm{i}S}\sum_{\{\sigma_i\}}\psi(\{\sigma_i\})\prod_{i=1}^N c^\dagger_{i;\sigma_i}\ket{0}\\
    \approx&\left(1-\mathrm{i}S\right)\sum_{\{\sigma_i\}}\psi(\{\sigma_i\})\prod_{i=1}^N c^\dagger_{i;\sigma_i}\ket{0}.
\end{split}
\end{equation}
Since there is no double occupancy and zero occupancy in Eq.~\eqref{eqGSphiinfty}, we have that $T_{-1}=0$ and $T_1$ becomes a free fermion hopping as 
$$T_1=-\sum_{\braket{ij},\sigma}(t_{ij}c^\dagger_{i;\sigma}c_{j;\sigma}+\mathrm{H.c.}).$$ Therefore, Eq.~\eqref{eqGSHubbard} can be simplified as
\begin{equation}\label{eqGSUfinite}
\begin{split}
\ket{\mathrm{GS},\mathrm{Hubbard}}
=&\left(1+\sum_{i,j,\sigma}\frac{t_{ij}}{U}c^\dagger_{i;\sigma}c_{j;\sigma}\right)\times\\
&\sum_{\{\sigma_i\}}\psi(\{\sigma_i\})\prod_{i=1}^N c^\dagger_{i;\sigma_i}\ket{0}.    
\end{split}
\end{equation}

The above equation can be represented as the ancilla wavefunction at large $\Phi$. 
The hopping terms of $c_{\sigma}$ and $f_{1\sigma}$ in $H^M_e$ in Eq. (2) in the main text are parameterized by $t_{c,ij}$ and $t_{1,ij}$, respectively. And their Fourier transform are denoted by $h_c(\mathbf{k})$ and $h_{f_1}(\mathbf{k})$, respectively. 
Then, the mean field Hamiltonian of $c_{\sigma}$ and $f_{1\sigma}$ can be written as:
\begin{equation}
\begin{split}
H^M_{e}=&\sum_{\mathbf{k},\sigma} \begin{pmatrix}
    c^\dagger_{\mathbf{k};\sigma} & f^\dagger_{\mathbf{k};1\sigma}
    \end{pmatrix}
h^M_e(\mathbf{k})
    \begin{pmatrix}
        c_{\mathbf{k};\sigma} \\
        f_{\mathbf{k};1\sigma}
    \end{pmatrix}\\
    =&\sum_{\mathbf{k},\sigma}
    \begin{pmatrix}
    c^\dagger_{\mathbf{k};\sigma} & f^\dagger_{\mathbf{k};1\sigma}
    \end{pmatrix}
    \begin{pmatrix}
        -h_c(\mathbf{k}) & \Phi \\
        \Phi & h_{f_1}(\mathbf{k}) 
    \end{pmatrix}
    \begin{pmatrix}
        c_{\mathbf{k};\sigma} \\
        f_{\mathbf{k};1\sigma}
    \end{pmatrix}\\
    =&\sum_{\mathbf{k},\sigma}
    \begin{pmatrix}
    c^\dagger_{\mathbf{k};\sigma} & f^\dagger_{\mathbf{k};1\sigma}
    \end{pmatrix}
    \tilde{h}^M_e
    \begin{pmatrix}
        c_{\mathbf{k};\sigma} \\
        f_{\mathbf{k};1\sigma}
    \end{pmatrix},
\end{split}
\label{EqHMF}
\end{equation}
with $h_e^M(\mathbf{k})$ the Fourier transform of $H^M_e$. Here, 
\begin{equation}
\begin{split}
&\tilde{h}^M_e =
    \frac{-h_c(\mathbf{k})+h_{f_1}(\mathbf{k})}{2}+
   e^{-\mathrm{i}\tilde S_\mathbf{k}}
     \begin{pmatrix}
        0 & \Phi^\prime(\mathbf{k}) \\
        \Phi^\prime(\mathbf{k}) & 0 
    \end{pmatrix}
    e^{\mathrm{i}\tilde S_\mathbf{k}},\\
&\Phi^\prime(\mathbf{k})=\sqrt{\Phi^2+\left(\frac{h_c(\mathbf{k})+h_{f_1}(\mathbf{k})}{2}\right)^2},\\
&\tilde S_\mathbf{k}=\frac{1}{2}\arctan\left(\frac{h_c(\mathbf{k})+h_{f_1}(\mathbf{k})}{2\Phi}\right)
    \begin{pmatrix}
        0 & -\mathrm{i} \\
        \mathrm{i} & 0
    \end{pmatrix}.
\end{split}
\end{equation}
By defining
\begin{equation}
    \tilde S = \sum_{\mathbf{k},\sigma} 
    \begin{pmatrix}
    c^\dagger_{\mathbf{k};\sigma} & f^\dagger_{\mathbf{k};1\sigma}
    \end{pmatrix}
\tilde S_\mathbf{k}
    \begin{pmatrix}
        c_{\mathbf{k};\sigma} \\
        f_{\mathbf{k};1\sigma}
    \end{pmatrix},
\end{equation}
for a given value of $\Phi$, we obtain
\begin{equation*}
\begin{split}
    H^M_e=&e^{-\mathrm{i}\tilde S}H^M_e(\Phi\rightarrow+\infty)e^{\mathrm{i}\tilde S}\\
    &+\sum_{\mathbf{k},\sigma}\frac{-h_c(\mathbf{k})+h_{f_1}(\mathbf{k})}{2}(c^\dagger_{\mathbf{k};\sigma}c_{\mathbf{k};\sigma}+f^\dagger_{\mathbf{k};1\sigma}f_{\mathbf{k};1\sigma}).    
\end{split}
\end{equation*} 
Note that the second term will not affect the ground state of $H^M_e$, as long as $\Phi$ is large enough to ensure that \textbf{there is exactly one eigenvalue of $h^M_e(\mathbf{k})$ less than $0$ (Fermi level) for each $\mathbf{k}$}. 
Therefore, $\ket{\mathrm{Slater}[c,f_1]}$ determined by $H^M_e$ can be obtained by performing a unitary transformation on Eq.~\eqref{eqGS1} as
\begin{equation}
    \begin{split}
        \ket{\mathrm{Slater}[c,f_1]}&=  e^{-\mathrm{i}\tilde{S}}\prod_{i=1}^N\left(\frac{1}{2}(c^\dagger_{i;\uparrow}-f^\dagger_{i;1\uparrow})(c^\dagger_{i;\downarrow}-f^\dagger_{i;1\downarrow})\right)\ket{0}\\
        \approx  (&1-\mathrm{i}\tilde S)\prod_{i=1}^N\left(\frac{1}{2}(c^\dagger_{i;\uparrow}-f^\dagger_{i;1\uparrow})(c^\dagger_{i;\downarrow}-f^\dagger_{i;1\downarrow})\right)\ket{0}.
    \end{split}
\end{equation}
By noting that 
\begin{equation}
\begin{split}
    1-\mathrm{i}\tilde S
   \approx &1-\sum_{\mathbf{k},\sigma}\frac{h_c(\mathbf{k})+h_{f_1}(\mathbf{k})}{4\Phi}(c^\dagger_{\mathbf{k};\sigma}f_{\mathbf{k};1\sigma}-f^\dagger_{\mathbf{k};1\sigma}c_{\mathbf{k};\sigma})\\
    =&1-\sum_{i,j,\sigma}\frac{t_{c,ij}+t_{1,ij}}{4\Phi}(c^\dagger_{i;\sigma}f_{j;1\sigma}-f^\dagger_{i;1\sigma}c_{j;\sigma}),
\end{split}
\end{equation}
we can find that the ancilla wavefunction doesn't depend on $t_{c,ij}$ and $t_{1,ij}$ individually, but rather the sum of them. Thus, we can assume that $t_{c,ij}=t_{ij}$ and $t_{1,ij}=0$ for simplicity. Using the relation in Eq.~\eqref{eqcdf}, the final projected state can be calculated as:
\begin{equation}\label{eqGSphifinite}
\begin{split}
    \ket{\Psi_c}&=\left(1+\sum_{i,j,\sigma}\frac{t_{ij}}{2\Phi}c^\dagger_{i;\sigma}c_{j;\sigma}\right)\times{}\\
    &~~\sum_{\{\sigma_i\}}\psi(\{\sigma_i\})\prod_{i=1}^N c^\dagger_{i;\sigma_i}\ket{0},
\end{split}
\end{equation}
which is the same as Eq.~\eqref{eqGSUfinite} if $\Phi=\frac{U}{2}$.
Eventually, we prove that for large $U$, this wavefunction is equivalent to applying an inverse Schiffer-Wolff transformation to the Gutzwiller projected wavefunction.

\begin{figure}[!t]
\includegraphics[width=1.\linewidth]{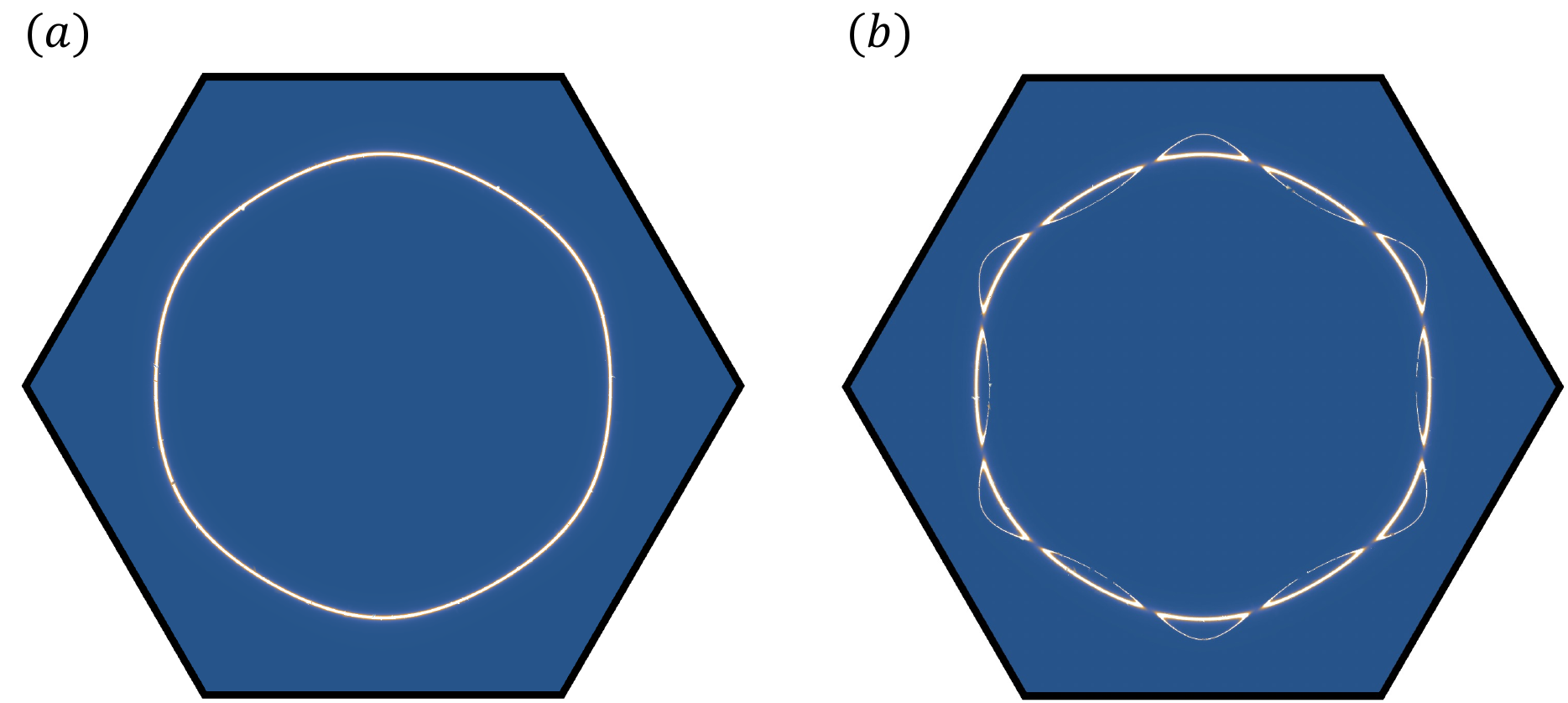}
\caption{The contour plots of the spectral weight $A_{f_1}(\mathbf{k},\omega=0)$. (a) $\Phi=0$. (b) $\Phi=0.1$. }
\label{fig:spectralf1}
\end{figure}

\begin{figure*}[!t]
\includegraphics[width=.8\linewidth]{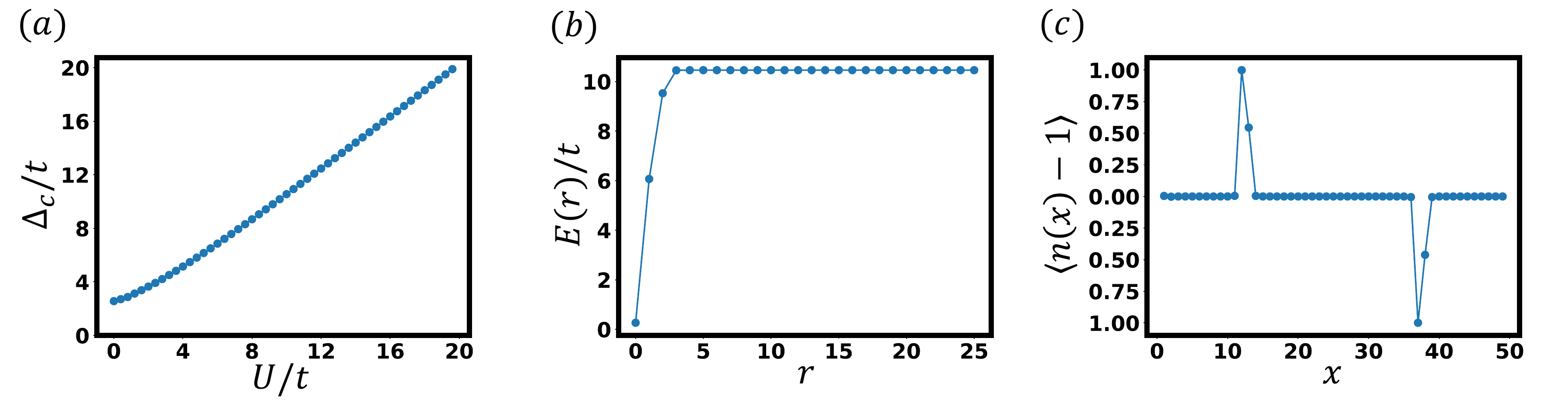}
\caption{(a) The charge gap of the ancilla wavefunction as a function of $U/t$. In (b)(c), we choose $\Phi/t=5$ and $U/t=9.9$. (b) The energy of an excited state versus the distance between the doublon and holon. Here the positions of doublon and holon are $r_1=\frac{L}{4}$ and $r_2=r_1+r$, respectively. (c) The local electron density for the excited state with $r_1=\frac{L}{4}$ and $r_2=\frac{3L}{4}$. Here we use $L=50$.}
\label{fig:excited_state}
\end{figure*}

\section{Calculation of the spectral weight function}~\label{app:spectral}

In this section, we show the details for calculating the spectral weight presented in Fig.4 in the main text. The spectral weight function of $c_{\sigma}$ is defined as~\cite{coleman2015introduction}:
\begin{equation}
    A_c(\mathbf{k},\omega)=\frac{1}{\pi}\mathrm{Im}G_c(\mathbf{k},\omega-\mathrm{i}\delta),
\end{equation}
where $G_c(\mathbf{k},\omega)$ is the Green's function of $c_{\sigma}$ and $\delta\rightarrow0^+$. The Green's function in the free fermion system can be written as~\cite{coleman2015introduction}:
\begin{equation}
G_c(\mathbf{k},\omega)=\sum_{\lambda}\frac{|M_{c,\lambda}(\mathbf{k})|^2}{\omega-\epsilon_\lambda(\mathbf{k})+\mathrm{i}\delta\mathrm{sgn}(\epsilon_{\lambda})},
\end{equation}
where $\epsilon_\lambda(\mathbf{k})$  ($\lambda=1,2$) is the eigenvalue of the Hamiltonian $h^M_e(\mathbf{k})$. The corresponding normalized eigenvector is denoted by $(M_{c,\lambda}(\mathbf{k}),M_{f_1,\lambda}(\mathbf{k}))^T$. The spectral weight of $f_1$ is defined similarly.
In our calculation, we choose a small value of $\delta=0.005$. The spectral weight of $f_{1\sigma}$ for the same Hamiltonian $h^M_e(\mathbf{k})$ is presented in Fig.~\ref{fig:spectralf1}.

\begin{figure}[!b]
\includegraphics[width=1\linewidth]{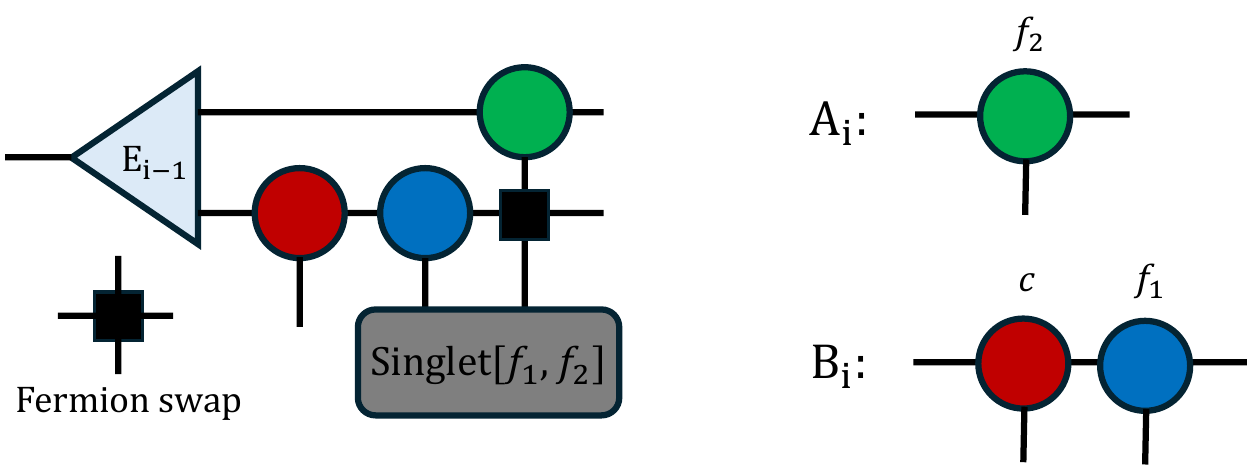}
\caption{The sketch of the tensor contraction.}
\label{fig:algrothim}
\end{figure}

\section{Excited state and charge gap of the ancilla wavefunction}~\label{app:Excited}

In this section, we define the charge gap of the ancilla wavefunction and show its dependence on $U/t$. The excited state with a doublon-holon pair in the ancilla wavefunction $\ket{\Psi_c}$ is introduced as 
\begin{equation}
P_S\sum_\sigma\psi^\dagger_{c;\sigma}(r_1)\psi_{v;\sigma}(r_2)\lvert\Psi_0\rangle,    ~\label{eq:cexcit}
\end{equation}
where $\psi_{c;\sigma}(r)$ and $\psi_{v;\sigma}(r)$ are the Wannier orbitals~\cite{Tu2020} of the conduction and valence bands of the Hamiltonian $H^M_e$ in Eq.~\eqref{EqHMF}, respectively. The distance between doublon and holon is defined as $r=|r_1-r_2|$.

In the thermodynamic limit, the energy of the excited state depends only on $r$ due to translational invariance, represented as $E(r)$. The charge gap is defined as $E(r\rightarrow+\infty)-E_0$, where $E_0$ is the energy of the ground state $\ket{\Psi_c}$. In our calculation, for a finite system of size $L$, we choose $r_1=\frac{L}{4}$ and  $r_2=\frac{3L}{4}$ to calculate the charge gap. Its dependence on $U/t$ is shown in Fig.~\ref{fig:excited_state}(a). Additionally, we select a special case of $\Phi/t=5$ and plot the energy $E(r)$ and local electron density $n(x)$ in Fig.~\ref{fig:excited_state}(b) and (c), respectively. 
We find that the final charge gap is roughly just the band gap of the mean-field $H^M_e$. This supports our intuition that the electrons $c_{\sigma}$, together with the ancilla fermions $f_{1\sigma}$, form the upper and lower Hubbard bands.
We note that the concept of the upper and lower Hubbard bands has existed for a long time. However, to the best of our knowledge, this is the first time they have been explicitly constructed within a simple mean-field parton theory.

\section{Details of the tensor network calculation}

In this section, we provide the details on the MPS construction of our ancilla wavefunction. 

For the system with spin SU(2) symmetry, we construct the correlation matrices~\cite{PhysRevB.92.075132} for $\ket{{\rm Slater}[c,f_1]}$ in the spin up and down channel, respectively. For a system with size $L$, each channel's correlation matrix $\Lambda$ is of $2L\times2L$. 
Consider a subsystem composed of $B$ fermions, corresponding to sites $1,2,...,B$. We can extract a submatrix $\mathcal{B}$ of correlation matrice with size $B\times B$, with the sites being restricted to $1,2,...,B$. 
The eigenvectors of $\mathcal{B}$ with eigenvalues closest to 1 or 0 are denoted as $v = (v_1, v_2, \dots, v_B)^T$.
We define a product of gates as $V_\mathcal{B}=V(\theta_{B-1})...V(\theta_2)V(\theta_1)$, where 
\begin{equation}
    V(\theta_i)=
    \begin{pmatrix}
        \cos\theta_i & -\sin\theta_i \\
        \sin\theta_i & \cos\theta_i
    \end{pmatrix}
\end{equation}
and $\theta_i=\arctan\frac{v_B}{v_{B-1}}$. By applying $V_\mathcal{B}$ to $\Lambda$, we will have $V^\dagger_\mathcal{B}\Lambda V_{\mathcal{B}}$ with $n_1=1$ or 0 as the top left entry. We let $\mathcal{B}_1=\mathcal{B}$ and perform the same procedure for subsystem composing sites $2,3,...,B+1$. Eventually we can get a diagonalized matrix $V^\dagger_{\mathcal{B}_{2L-1}}...V^\dagger_{\mathcal{B}_2}V^\dagger_{\mathcal{B}_1}\Lambda V_{\mathcal{B}_1} V_{\mathcal{B}_2}...V_{\mathcal{B}_{2L-1}}$. The eigenvalue $n_i$ is $0$ or $1$. We construct the fermionic gaussian state by applying $V_{\mathcal{B}_1} V_{\mathcal{B}_2}...V_{\mathcal{B}_{2L-1}}$ to a product state $\otimes_{i=1}^{2L}\lvert{n_i}\rangle$. Here we assume $\lvert0\rangle$ is the unoccupied state and $\lvert1\rangle$ is the occupied state. The state $\lvert \mathrm{Gauss}[f_2]\rangle$ can be constructed in a similar manner.

Finally, we use the Gutzwiller zipper method~\cite{aghaei2020efficient} to perform a tensor product of $\lvert\mathrm{Slater}[c,f_1]\rangle$ and $\lvert\mathrm{Gauss}[f_2]\rangle$, then implement the desired singlet projection $P_S$. The method is to perform tensor product and projection site by site. For site $i$, we have an input $3$-leg tensor $E_{i-1}$ from previous sites (for site $i=1$, $E_0=I$). Meanwhile, we have $A_i$ and $B_i$ as the right-canonical form~\cite{schollwock2011} of $\lvert\mathrm{Slater}[c,f_1]\rangle$ and $\lvert\mathrm{Gauss}[f_2]\rangle$, respectively. As shown in Fig.~\ref{fig:algrothim}, after performing the tensor contraction, a new tensor $C_i$ for the ancilla wavefunction is eventually obtained. 

\section{Additional details on the numerical results}~\label{app:results}

In this section, we provide more details about the numerical results of energy and overlap.

\begin{figure}[!t]
\includegraphics[width=1\linewidth]{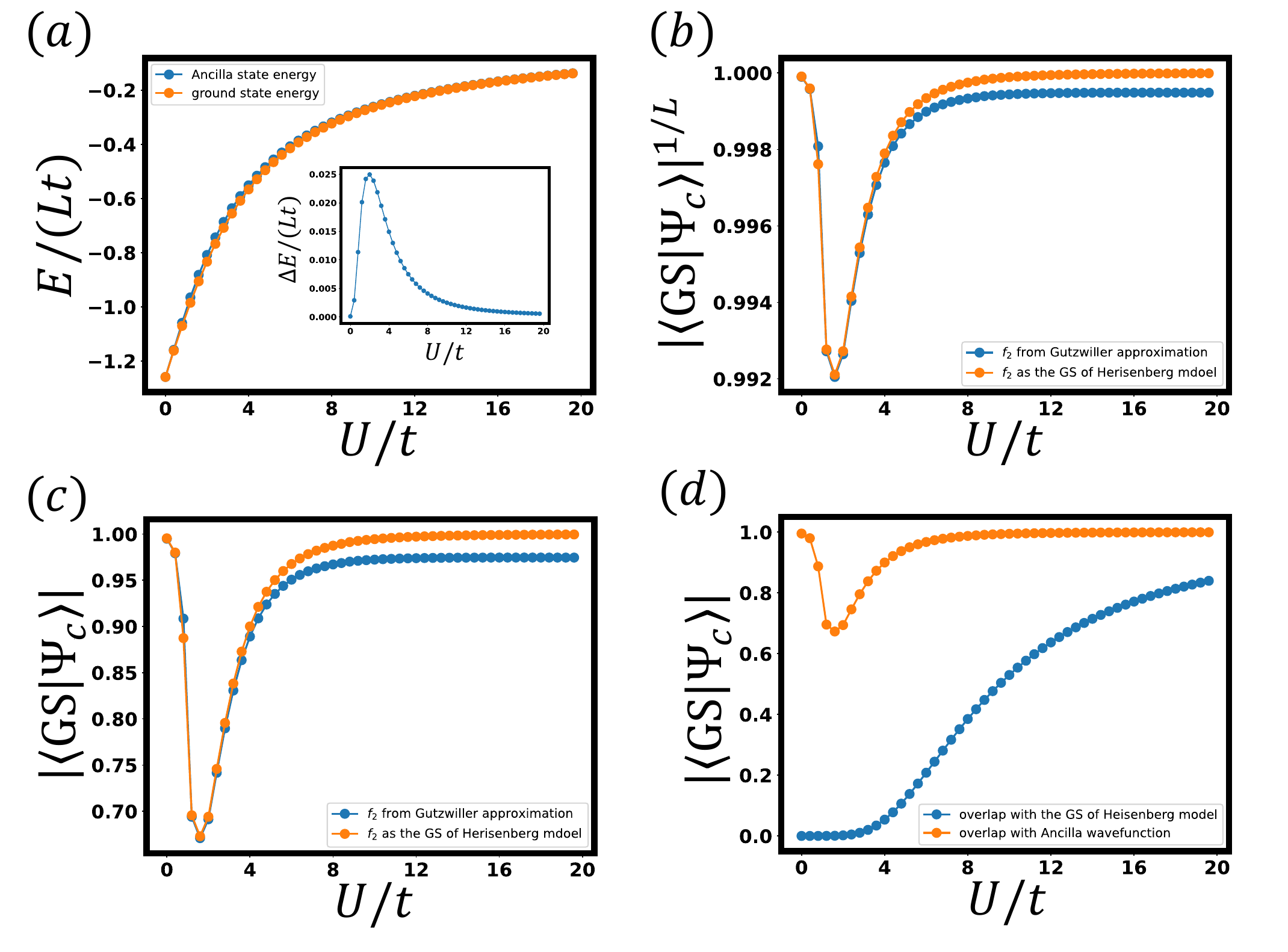}
\caption{ Benchmark of two ancilla wavefunctions prepared by two different $\ket{{\rm Gauss}[f_2]}$ ansatz, e.g., (i) $\ket{\rm GS}_S$ the ground state of the 1D antiferromagnetic Heisenberg chain obtained by DMRG and (ii) $\ket{\rm GFS}$ the Gutzwiller projected Fermi sea state on the 1D chain.
(a) The per-site energy for the ancilla wavefunction prepared with $\ket{\rm GS}_S$ and the DMRG energy as functions of $U/t$. The inset represents the energy difference. (b) The per-site overlap between the ancilla wavefunction and $|{\rm GS}\rangle$ (the ground state for the 1D Hubbard chain obtained by DMRG). 
(c) shows the overlap for the entire system.
(d) The overlap between $|{\rm GS}\rangle$ and (i) the ancilla wavefunction prepared with $|{\rm GS}\rangle_S$ as well as (ii) $|{\rm GS}_S\rangle$. Here we set $L=50$ and the bond dimension $\chi=364$. }
\label{fig:overlap_more}
\end{figure}

\begin{figure}[!b]
    \includegraphics[width=1.\linewidth]{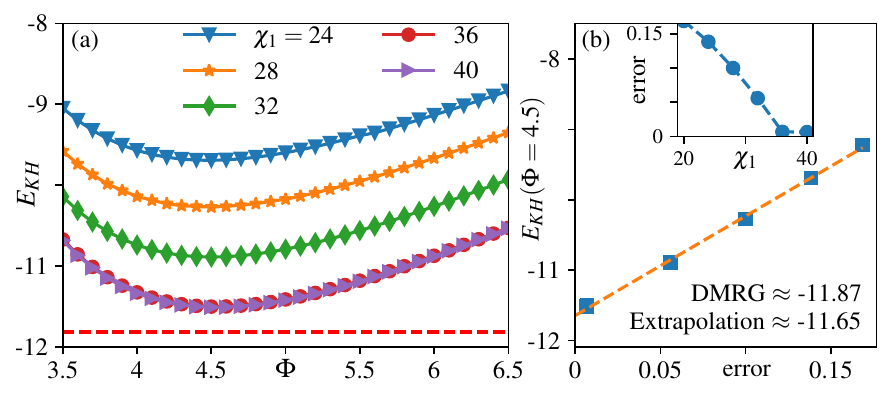}
    \caption{ The variational energy $E_{\rm KH}$ by ancilla wavefunctions for the Kitaev-Hubbard model in Eq.~\eqref{eq:KHmodel}. (a) The best variational energy is obtained at $\Phi=U/2$, as predicted by our theory. (b) The extrapolation of the best variational energy to the accumulated truncation error of $|{\rm Slater}[c, f_{1}]\>$. Inset: Accumulated truncation error as a function of $\chi_{1}$. Here we focus on the case of $L_y=4$, $L_x=12$, and $U/t=9$.}
    \label{fig:FigKconKH}
\end{figure}

When calculating the 1D Hubbard chain using DMRG, the bond dimension of MPS is fixed as $\chi=364$, resulting in a tiny truncation error of $\epsilon\sim{}10^{-11}$. The same bond dimension has been used for the construction of ancilla wavefunction $\ket{\Psi_c}$. 
To verify the effects of $\ket{{\rm Gauss}[f_2]}$, we use two different $\ket{{\rm Gauss}[f_2]}$ states to construct our ancilla wavefunctions. The first one state $\ket{\rm GS}_S$, as mentioned in the main text, is the ground state of the 1D antiferromagnetic Heisenberg chain obtained by DMRG. We use $\ket{\rm GFS}$ the Gutzwiller projected Fermi sea state on the 1D chain as the second candidate. The results are summarized in Fig.~\ref{fig:overlap_more}(a-c). We find that $\ket{\rm GFS}$ also works well. Nevertheless, since $\ket{\rm GFS}$ is not the true ground state of the spin model, its performance is slightly worse than that of $\ket{\rm GS}_S$.
Furthermore, Fig.~\ref{fig:overlap_more}(d) illustrates the overlap between $\ket{\rm GS}_S$ and $\ket{\rm GS}$. The results indicate that the spin ground state $\ket{\rm GS}_S$ deviates significantly from the ground state of the Hubbard model at finite but large $U/t$ because of the emergence of the doublons and holons.  On the other hand, our ancilla wavefunction works quite well in capturing these charge fluctuations. 

For the 2D Kitaev-Hubbard model, the bond dimension for preparing the Kiteav QSL is $\chi_{2}=180$ (with an accumulated truncation error of $\approx0.02$). The DMRG state, which is initialized with our ancilla wavefunction, is converged at bond dimension $\chi_{D}=8000$ with typical truncation error $\sim{}10^{-6}$. We vary the bond dimension $\chi_1$ to examine the convergence of constructing the ancilla wavefunctions. As shown in Fig.~\ref{fig:FigKconKH}, the calculation converges with a bond dimension of $\chi_1\sim{}40$ for relatively large $U$.

\end{document}